\documentclass[lettersize,journal]{IEEEtran}
\usepackage{amsmath,amsfonts}
\usepackage{algorithmic}
\usepackage{algorithm}
\usepackage{array}
\usepackage[caption=false,font=normalsize,labelfont=sf,textfont=sf]{subfig}
\usepackage{textcomp}
\usepackage{stfloats}
\usepackage{url}
\usepackage{verbatim}
\usepackage{graphicx}
\usepackage{cite}
\usepackage{float}
\usepackage{bm}
\usepackage{amssymb}
\usepackage{stfloats}
\pdfoutput=1
\begin{document}
	
	\title{Two-Timescale Design for AP Mode Selection of Cooperative ISAC Networks}
	
	\author{Zhichu Ren, Cunhua Pan, \IEEEmembership{Senior Member, IEEE}, Hong Ren, \IEEEmembership{Member, IEEE}, Dongming Wang, \IEEEmembership{Member, IEEE}, Lexi Xu, \IEEEmembership{Senior Member, IEEE}, Jiangzhou Wang, \IEEEmembership{Fellow, IEEE}
		% <-this % stops a space
%		\thanks{
%		The work of Hong Ren was supported in part by the National Natural Science Foundation of China under Grant 62101128, and in part by Basic Research Project of Jiangsu Provincial Department of Science and Technology under Grant BK20210205. The work of Cunhua Pan was supported in part by the National Natural Science Foundation of China under Grants 62201137 and 62331023. The work of Cunhua Pan and Hong Ren was supported in part by the Fundamental Research Funds for the Central Universities under Grant 2242022k60001.}
		% <-this % stops a space
		\thanks{Z. Ren, C. Pan, H. Ren, D. Wang, and J. Wang are with National Mobile Communications Research Laboratory, Southeast University, Nanjing, China. (e-mail:\{220241017, cpan, hren, wangdm, j.z.wang\}@seu.edu.cn). L. Xu is with the China Unicom Research Institute, Beijing 100048, China.(e-mail: davidlexi@hotmail.com).
	
			\emph{Corresponding authors: Cunhua Pan}}}
	
	\markboth{}
	{Shell \MakeLowercase{\textit{et al.}}: A Sample Article Using IEEEtran.cls for IEEE Journals}
	
	
	\maketitle
	
	\begin{abstract}
		As an emerging technology, cooperative bi-static integrated sensing and communication (ISAC) is promising to achieve high-precision sensing, high-rate communication as well as self-interference (SI) avoidance. This paper investigates the two-timescale design for access point (AP) mode selection to realize the full potential of the cooperative bi-static ISAC network with low system overhead, where the beamforming at the APs is adapted to the rapidly-changing instantaneous channel state information (CSI), while the AP mode is adapted to the slowly-changing statistical CSI. We first apply the minimum mean square error (MMSE) estimator to estimate the channel between the APs and the channels from the APs to the user equipments (UEs). Then we adopt the low-complexity maximum ratio transmission (MRT) beamforming and the maximum ratio combining (MRC) detector, and derive the closed-form expressions of the communication rate and the sensing signal-to-interference-plus-noise-ratio (SINR). We formulate a non-convex integer optimization problem to maximize the minimum sensing SINR under the communication quality of service (QoS) constraints. McCormick envelope relaxation and successive convex approximation (SCA) techniques are applied to solve the challenging non-convex integer optimization problem. Simulation results validate the closed-form expressions and prove the convergence and effectiveness of the proposed AP mode selection scheme.
	\end{abstract}
	
	\begin{IEEEkeywords}
		cooperative integrated sensing and communication, two-timescale design, access point mode selection, integer optimization.
	\end{IEEEkeywords}
	
	\section{Introduction}
	\IEEEPARstart{W}{ith} the advancement of researches on sixth-generation (6G) mobile communication systems, integrated sensing and communication (ISAC) has  been envisioned as a key technology, which is expected to be applied in various scenarios such as enhanced localization, vehicle-to-everything (V2X), and Internet-of-Things (IoT) \cite{liu2022integrated}. In recent years, extensive researches have been carried out in ISAC, such as the trade-off between communication and sensing performance, waveform design, and signal processing techniques \cite{wei2023integrated}.\\
	\indent Based on different modes of sensing, the architecture of ISAC is mainly categorized as mono-static and bi-static \cite{liu2022survey}. As a simpler architecture, mono-staic ISAC has received substantial attention. The authors of \cite{lu2022degrees} investigated the degrees of freedom (DoFs) achieved in mono-static ISAC systems. Training and transmission design schemes were developed to minimize the mean-squared errors (MSEs) of data transmission and target estimation in mono-static ISAC systems under typical block-fading channels \cite{he2024mse}. In mono-static ISAC systems, the base stations (BSs) have to operate in full-duplex (FD) mode, resulting in severe self-interference (SI) \cite{tang2024interference}. To tackle this issue, a novel waveform design strategy was proposed to mitigate SI while simultaneously increasing the communication rate in \cite{xiao2022waveform}. In contrast, bi-static ISAC offers the distinct advantage of eliminating SI while obtaining more location information of the targets. An et al. \cite{an2023fundamental} characterized the fundamental performance trade-off between the detection probability and the achievable rate in bi-static ISAC systems. A general framework was proposed for the joint target-localization and data detection in bi-static ISAC systems by exploiting both deterministic known and random unknown symbols \cite{zhao2024joint}. The authors of \cite{bauhofer2024bistatic} presented a maximum likelihood (ML) framework for adaptive fusion of bi-static measurements, including transmitter angle, receiver angle, and bi-static range to enhance sensing accuracy. \\
	\indent Nevertheless, it fails to fulfill the potential of ISAC by using the mono-static architecture with only a single base station or the bi-static architecture with only a pair of transceivers. In contrast, cooperative ISAC with multiple BSs or access points (APs) is expected to achieve long-range and precise sensing in \cite{wei2023integrated2}, \cite{xie2023collaborative}, and \cite{cui2024integrated}. A number of works have studied the  cooperative ISAC with multiple APs. In order to enhance system performance, several beamforming design methods were proposed in \cite{yang2024coordinated}, \cite{liu2024joint}, and \cite{babu2024precoding}. The performance of cooperative ISAC networks, including communication ergodic rates, Cram\'{e}r-Rao lower bound (CRLB), and asymptotic outage probabilities, were analyzed, and a resource allocation scheme was designed in \cite{liu2022performance} and \cite{zeng2023integrated}. Zhang et al. \cite{zhang2024target} proposed a cooperative ISAC framework by leveraging information-bearing orthogonal frequency division multiplexing (OFDM) \cite{zhu2009chunk}, \cite{zhu2011chunk} signals to attain the cooperative gain. Furthermore, it is inevitable that the cooperative network-level ISAC will result in an increase in mutual interference. The authors of \cite{xu2024interference} investigated the coordinated cellular network-based and distributed
	antenna-based ISAC systems from an interference management perspective and discussed associated optimization methods. \\
	\indent However, the position and function of APs are relatively fixed in the aforementioned works, thus failing to utilize the additional DoFs and spatial diversity gain brought about by multiple APs. It is therefore meaningful to study the methods used to select the downlink (DL) and uplink (UL) modes for the APs. Xu et al. \cite{xu2023integrated} proposed a heuristic AP mode selection algorithm based on the distance from the APs to the user equipments (UEs) and the  targets. The AP mode selection and beamforming were jointly optimized to achieve the trade-off between communication and sensing performance in \cite{liu2024cooperative} and \cite{xu2023joint}.\\
	\indent Nevertheless, the aforementioned works both designed the AP mode selection based on the instantaneous channel state information (CSI). In this way, the mode of APs needs to be frequently adjusted in each channel coherence interval, leading to extremely high computational complexity and system overhead. To address these practical challenges, a promising solution named the two-timescale scheme was recently proposed. In the two-timescale scheme, the AP beamforming is designed based on the instantaneous CSI, while the AP mode selection is optimized based on the long-term statistical CSI, such as the location of the UEs and the targets with respect to the APs. Recently, a few works have studied the AP mode selection of cooperative ISAC networks based on the statistical CSI. The authors of \cite{elfiatoure2024multiple} derived the closed-form expressions for the DL spectral efﬁciency (SE) of the UEs and the mainlobe-to-average-sidelobe ratio (MASR) of the sensing zones and jointly optimized the mode of APs and power control. Zeng et al. \cite{zeng2024multi} derived the closed-form expressions for the comunication rate and the localization error rate under imperfect CSI, and proposed a deep Q-network-based AP mode selection optimization algorithm. However, the researches mentioned above are not sufficiently comprehensive. Firstly, the authors of \cite{elfiatoure2024multiple} considered a cooperative ISAC network without UL UEs, and failed to consider the cross-link interference comprehensively. Secondly, reinforcement learning techniques were adopted to optimize the operation mode of APs in \cite{zeng2024multi}, which lacks interpretability. Ultimately, both authors of \cite{elfiatoure2024multiple} and \cite{zeng2024multi} considered Rayleigh channel model between the APs and the UEs, taking only the none-line-of-sight (NLoS) components, which is not general enough. Indeed, the line-of-sight (LoS) components need to be analyzed.\\
	\begin{figure}[t]
		\centering
		\includegraphics[width=3.5in]{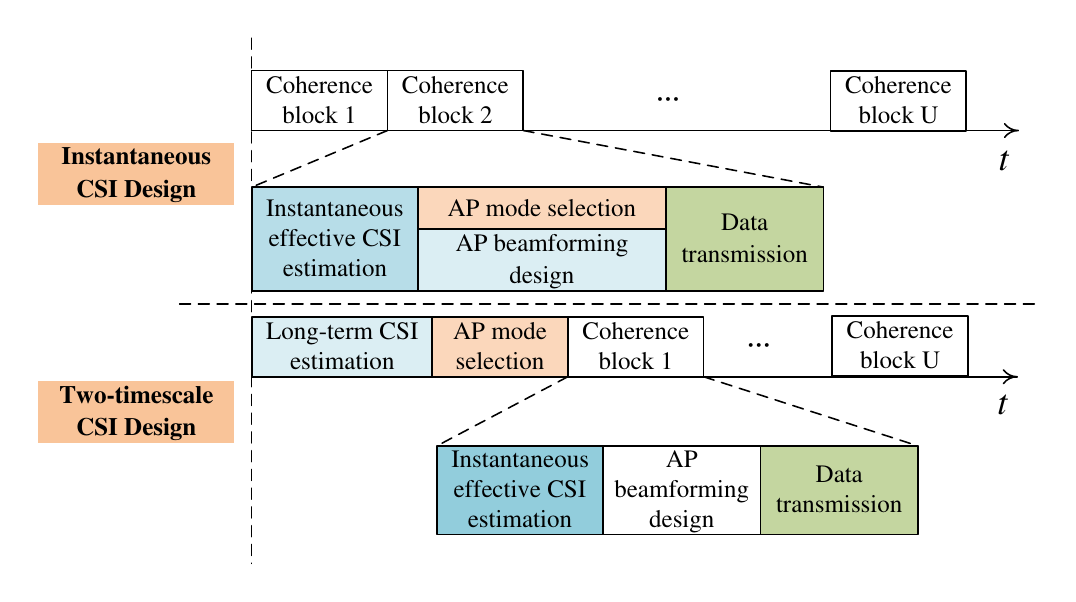}
		\caption{Illustration of AP mode selection designs based on instantaneous CSI and long-term statistical CSI ($U$ coherence blocks).}
		\label{fig1}
	\end{figure}
	\indent Motivated by the above background, our main contributions are given as follows:
	\begin{enumerate}
		\item We propose to apply the two-timescale scheme to the cooperative bi-static ISAC network for the AP mode selection design, which is shown in Fig. \ref{fig1}. The channels between the APs and the channels from the APs to the UEs are estimated by the minimum mean square error (MMSE) estimator to design the maximum ratio transmission (MRT) beamforming and the maximum ratio combining (MRC) detector. We also derive the closed-form expressions of the communication rate and the sensing signal-to-interference-plus-noise-ratio (SINR).
		\item Based on the analytical results of communication and sensing performance, we formulate a non-convex integer optimization problem to maximize the minimum sensing SINR under the communication QoS constraints and adopt McCormick envelope relaxation and successive convex approximation (SCA) techniques to solve this NP-hard problem. 
		\item Simulation results verify the accuracy of the closed-form expressions and demonstrate the superior advantages of the proposed AP mode selection scheme over other benchmark schemes. Furthermore, the two-timescale design provides a practical AP mode selection strategy for the cooperative bi-static ISAC network with extremely low system overhead.
	\end{enumerate}
	
	\indent The remaining sections are organized as follows. The system model is provided in Section \ref{System Model}. Then, the related channels are estimated in Section \ref{Channel Estimation}. In Section \ref{Analysis}, the closed-form expressions of the communication rate and the sensing SINR are derived. In Section \ref{AP Mode Selection Algorithm}, the AP mode selection is optimized. In Section \ref{simulation}, extensive numerical results are presented. Finally, our conclusion is drawn in Section \ref{conclusion}.\\
%	\indent \emph{Notations}: ${{\mathbb{ C}}^{M \times N}}$ and ${{\mathbb{ R}}^{M \times N}}$ denote the set of $M \times N$ complex and real matrices, respectively. ${{\mathrm{Re}}}\{\cdot\} $, ${{\mathbb{E}}}\{\cdot\} $ and ${{\mathrm{Cov}}}\{\cdot\} $ denote the real, expectation, and covariance operators, respectively. ${\left\| {\bf{x}} \right\|_2}$ denotes the 2-norm of vector ${\bf{x}}$. $|a|$ denotes the absolute value of a complex number $a$. ${\rm{Tr}}\left( {\bf{A}} \right)$ denotes the trace operation of ${\bf{A}}$. ${\cal C}{\cal N}({\bf{0}},{\bf{I}})$ represents a random vector following the distribution of zero mean and unit variance matrix. ${\left( \cdot \right)^{-1}}$ and ${\left(  \cdot \right)^{\rm{H}}}$ denote the inverse and Hermitian operators, respectively. $\mathcal{O}$ denotes the standard big-O notation.
	\section{System Model}\label{System Model}
	\begin{figure}[h]
		\centering
		\includegraphics[width=3.5in]{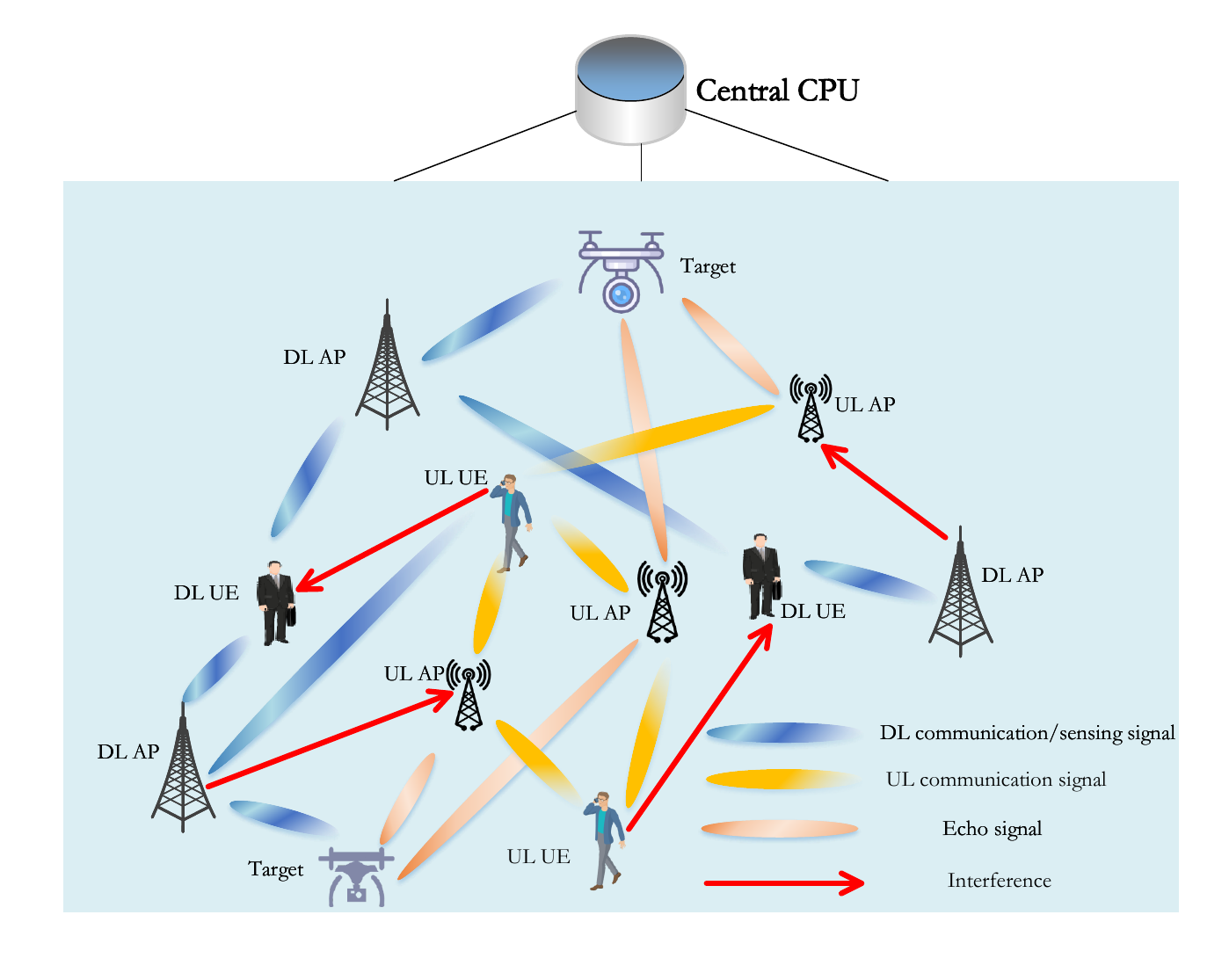}
		\caption{System model for a cooperative bi-static ISAC network.}
		\label{fig2}
	\end{figure}
	\indent We consider a cooperative ISAC network with bi-static architecture under TDD operation, which simultaneously enables DL communication, UL data transmission and target sensing. As illustrated in Fig. \ref{fig2}, the network is comprised of a central processing unit (CPU), $M$ APs with $N$ antennas, $K_{\mathrm{dl}}$ DL UEs with single antennas, $K_{\mathrm{ul}}$ UL UEs with single antennas and $K_{\mathrm{t}}$ passive targets to be sensed. All the APs and UEs function in the half-duplex mode. Let $\mathcal{M}_{\mathrm{dl}}=\{1,\dots,M_{\mathrm{dl}}\}$, $\mathcal{M}_{\mathrm{ul}}=\{1,\dots,M_{\mathrm{ul}}\}$, $\mathcal{K}_{\mathrm{dl}}=\{1,\dots,K_{\mathrm{dl}}\}$, $\mathcal{K}_{\mathrm{ul}}=\{1,\dots,K_{\mathrm{ul}}\}$, $\mathcal{K}_{\mathrm{t}}=\{1,\dots,K_{\mathrm{t}}\}$ denote the set of DL APs, UL APs, DL UEs, UL UEs and passive targets to be sensed.\\
	\indent The channel between the $m$-th AP and the $k$-th UE is modeled as Rician fading channel which can be expressed as 
	\begin{equation}
		\mathbf{g}_{m,k}=\sqrt{\frac{\beta_{m,k}}{\delta_{m,k}+1}}(\sqrt{\delta_{m,k}}\overline{\mathbf{g}}_{m,k}+\widetilde{\mathbf{g}}_{m,k}),
	\end{equation}
	where $\beta_{m,k}$ represents the large-scale fading coefficient, $\delta_{m,k}$ represents the Rician factor, $\overline{\mathbf{g}}_{m,k}$ denotes the LoS component, and $\widetilde{\mathbf{g}}_{m,k}$ denotes the NLoS component of which elements are i.i.d. complex Gaussian random variables with zero mean and unit variance. For the LoS paths, the uniform planar array (UPA) model is adopted for the APs. Hence, $\overline{\mathbf{g}}_{m,k}$ is modelled as follows
	\begin{equation}
		\overline{\mathbf{g}}_{m,k}=\mathbf{a}(\theta^{\mathrm{c}}_{m,k},\phi^{\mathrm{c}}_{m,k}),
	\end{equation}
	where $\theta^{\mathrm{c}}_{m,k}$ and $\phi^{\mathrm{c}}_{m,k}$ denote the azimuth angle and the elevation angle between the $m$-th AP and the $k$-th UE, and $\mathbf{a}(\theta,\phi)$ denotes the steering vector which is given by
	\begin{equation}
		\begin{aligned}
			\mathbf{a}(\theta,\phi)&=[1,\cdots,e^{j\frac{2\pi d}{\lambda}(i_x\cos\theta\cos\phi+i_z\sin\theta)},\\&\cdots e^{j\frac{2\pi d}{\lambda}((N_x-1)\cos\theta\cos\phi+(N_z-1)\sin\theta)}]^{\mathrm{T}} ,
		\end{aligned}
	\end{equation}
	where $d$ and $\lambda$ denote the inter-antenna spacing and the signal wavelength, and $1\leq i_x\leq N_x-1$, $1\leq i_z\leq N_z-1$ denote the antenna indices in the $x-z$ plane.\\
	\indent The channel between the $i$-th UE and the $j$-th UE, and the channel between the $m$-th AP and the $l$-th AP are both modeled as Rayleigh fading channel which can be respectively expressed as
	\begin{equation}
		h_{i,j}=\sqrt{\alpha_{i,j}}\widetilde{h}_{i,j},
	\end{equation}
	\begin{equation}
		\mathbf{H}_{m,l}=\sqrt{\gamma_{m,l}}\widetilde{\mathbf{H}}_{m,l},
	\end{equation}
	where $\alpha_{i,j}$ and $\gamma_{m,l}$ denote the large-scale fading coefficients, both $\widetilde{h}_{i,j}$ and $\widetilde{\mathbf{H}}_{m,l}$ denote the NLoS components of which elements are i.i.d. complex Gaussian random variables with zero mean and unit variance.
	%\begin{table*}
	%	\caption{List of Main Symbols\label{tab:table1}}
	%	\centering
	%	\begin{tabular}{|c|c|}
		%		\hline
		%		$K_{\mathrm{dl}}(K_{\mathrm{ul}})$ & Number of DL (UL) UEs\\
		%		\hline
		%		$M_{\mathrm{c}}(M_{\mathrm{s}})$ & Number of communication (sensing) APs\\
		%		\hline
		%		$K_{\mathrm{t}}$ & Number of targets\\
		%		\hline
		%		$N$ & Number of antennas per-AP\\
		%		\hline
		%		$\mathbf{g}_{m,k}$ & Channel between UE $k$ and AP $m$\\
		%		\hline
		%		$\mathbf{H}_{m,l}$ & Channel between communication AP $m$ and sensing AP $l$\\
		%		\hline
		%		$h_{i,j}$ & Channel between UE $i$ and UE $j$\\
		%		\hline
		%		$\beta_{m,k}, \alpha_{i,j}, \gamma_{m,l}$ & Large scale fading coefficient\\
		%		\hline
		%		$\delta_{m,k}$ & Rician factor\\
		%		\hline
		%		$\tau_{c}(\tau_{\mathrm{dl}}/\tau_{\mathrm{ul}})$ & Lengths of coherence interval (pilot signal for DL/UL channel estimation)\\
		%		\hline
		%		$p_{\mathrm{ul}}(p_{\mathrm{dl}})$ & Transmit power of UL UEs (communication APs)\\
		%		\hline
		%		$\sigma^2$ & Power of thermal noise\\
		%		\hline
		%		$a_m (b_m)$ & Binary variables to indicate the mode assignment\\
		%		\hline
		%		$\boldsymbol{\varphi}_k(\boldsymbol{\phi}_m)$ & UL (DL) pilot\\
		%		\hline
		%	\end{tabular}
	%\end{table*}
	\section{Channel Estimation}\label{Channel Estimation}
	\subsection{Estimate the Channel Between the APs and the UEs}
	\indent In each coherence block of length $\tau_{c}$, all the UEs send mutually orthogonal UL pilot to the APs, and channel $\mathbf{g}_{m,k}$ is estimated by the received pilot signals and the MMSE estimator. The pilot sent by the $k$-th UE is denoted as $\boldsymbol{\varphi}_k \in \mathbb{C}^{\tau_{\mathrm{ul}} \times 1}$ which satisfies $\left\| \boldsymbol{\varphi}_k \right\|_2^2=1$ and $\tau_{\mathrm{ul}}\geq K_{\mathrm{dl}}+K_{\mathrm{ul}}$. The pilot signal received at the $m$-th AP can be expressed as
	\begin{equation}\label{1}
		\mathbf{Y}_{m}=\displaystyle\sum_{k=1}^{K_{\mathrm{dl}}+K_{\mathrm{ul}}}\sqrt{\tau_{\mathrm{ul}} p_{\mathrm{ul}}}\mathbf{g}_{m,k}\boldsymbol{\varphi}_k^{\mathrm{H}}+\mathbf{N}_m,
	\end{equation}
	where $\tau_{\mathrm{ul}} p_{\mathrm{ul}}$ denotes the transmit UL pilot power, and $\mathbf{N}_m\sim \mathcal{CN}(\mathbf{0},\sigma^2\mathbf{I}_N)$ denotes the additive white Gaussian noise (AWGN). Multiplying \eqref{1} by $\boldsymbol{\varphi}_k$ and utilizing the orthogonality of the pilot signals, the observation vector of the $k$-th UE at the $m$-th AP can be expressed as follows
	\begin{equation}
		\mathbf{y}_{m,k}=\sqrt{\tau_{\mathrm{ul}} p_{\mathrm{ul}}}\mathbf{g}_{m,k}+\mathbf{N}_m\boldsymbol{\varphi}_k.
	\end{equation}
	\indent Several auto-correlation matrices are calculated as follows
	\begin{equation}
		\begin{aligned}
			&\mathbf{R}_{\mathbf{y}_{m,k}\mathbf{y}_{m,k}}=\mathbb{E}\{\mathbf{y}_{m,k}\mathbf{y}_{m,k}^{\mathrm{H}}\}\\&=\frac{\tau_{\mathrm{ul}} p_{\mathrm{ul}} \beta_{m,k}\delta_{m,k}}{\delta_{m,k}+1}\overline{\mathbf{g}}_{m,k}\overline{\mathbf{g}}_{m,k}^{\mathrm{H}}+(\frac{\tau_{\mathrm{ul}} p_{\mathrm{ul}} \beta_{m,k}}{\delta_{m,k}+1}+\sigma^2)\mathbf{I}_{N},
		\end{aligned}
	\end{equation}
	\begin{equation}
		\begin{aligned}
			&\mathbf{R}_{\mathbf{g}_{m,k}\mathbf{y}_{m,k}}=\mathbb{E}\{\mathbf{g}_{m,k}\mathbf{y}_{m,k}^{\mathrm{H}}\}\\&=\frac{\sqrt{\tau_{\mathrm{ul}} p_{\mathrm{ul}}} \beta_{m,k}\delta_{m,k}}{\delta_{m,k}+1}\overline{\mathbf{g}}_{m,k}\overline{\mathbf{g}}_{m,k}^{\mathrm{H}}+\frac{\sqrt{\tau_{\mathrm{ul}} p_{\mathrm{ul}}} \beta_{m,k}}{\delta_{m,k}+1}\mathbf{I}_{N}.
		\end{aligned}
	\end{equation}
	\indent By defining $v_{m,k}=\frac{ \beta_{m,k}\delta_{m,k}}{\delta_{m,k}+1}$, $u_{m,k}=\frac{ \beta_{m,k}}{\delta_{m,k}+1}$ and using the Woodbury matrix identity \cite{kay1993fundamentals}, we have
	\begin{equation}
		\begin{aligned}
			&\mathbf{R}_{\mathbf{y}_{m,k}\mathbf{y}_{m,k}}^{-1}=(\tau_{\mathrm{ul}} p_{\mathrm{ul}}u_{m,k}+\sigma^2)^{-1}\mathbf{I}_{N}\\&-\frac{\tau_{\mathrm{ul}} p_{\mathrm{ul}}v_{m,k}(\tau_{\mathrm{ul}} p_{\mathrm{ul}}u_{m,k}+\sigma^2)^{-2}}{1+N\tau_{\mathrm{ul}} p_{\mathrm{ul}}v_{m,k}(\tau_{\mathrm{ul}} p_{\mathrm{ul}}u_{m,k}+\sigma^2)^{-1}}\overline{\mathbf{g}}_{m,k}\overline{\mathbf{g}}_{m,k}^{\mathrm{H}}.
		\end{aligned}
	\end{equation}
	\indent Thus, the MMSE estimation of $\mathbf{g}_{m,k}$ can be expressed as
	\begin{equation}
		\begin{aligned}
			\hat{\mathbf{g}}_{m,k}&=\mathbf{R}_{\mathbf{g}_{m,k}\mathbf{y}_{m,k}}\mathbf{R}_{\mathbf{y}_{m,k}\mathbf{y}_{m,k}}^{-1}\mathbf{y}_{m,k}\\&= \mathbf{A}_{m,k}\mathbf{y}_{m,k}=\sqrt{\tau_{\mathrm{ul}} p_{\mathrm{ul}}}\mathbf{A}_{m,k}\mathbf{g}_{m,k}+\mathbf{A}_{m,k}\mathbf{N}_m\boldsymbol{\varphi}_k,
		\end{aligned}
	\end{equation}
	where
	\begin{equation}
		\begin{aligned}	\mathbf{A}_{m,k}&=\frac{\sqrt{\tau_{\mathrm{ul}}p_{\mathrm{ul}}}\sigma^2v_{m,k}\overline{\mathbf{g}}_{m,k}\overline{\mathbf{g}}_{m,k}^{\mathrm{H}}}{(\tau_{\mathrm{ul}}p_{\mathrm{ul}}u_{m,k}+\sigma^2)(\tau_{\mathrm{ul}}p_{\mathrm{ul}}u_{m,k}+\sigma^2+N\tau_{\mathrm{ul}}p_{\mathrm{ul}}v_{m,k})}\\&+\frac{\sqrt{\tau_{\mathrm{ul}}p_{\mathrm{ul}}}u_{m,k}}{\tau_{\mathrm{ul}}p_{\mathrm{ul}}u_{m,k}+\sigma^2}\mathbf{I}_N.
		\end{aligned}
	\end{equation}
	\indent The estimation error of $\mathbf{g}_{m,k}$ is denoted as $	\mathbf{e}_{m,k}^{\mathrm{C}}=\mathbf{g}_{m,k}-\hat{\mathbf{g}}_{m,k}$, then the mean squared error (MSE) matrix of the estimated channel is calculated as follows
	\begin{equation}
		\begin{aligned}
			&\mathbf{MSE}_{\mathbf{g}_{m,k}}\\&=\mathbb{E}\{\mathbf{e}_{m,k}^{\mathrm{C}}(\mathbf{e}_{m,k}^{\mathrm{C}})^{\mathrm{H}}\}=\mathbb{E}\{\mathbf{g}_{m,k}\mathbf{g}_{m,k}^{\mathrm{H}}\}-\mathbb{E}\{\hat{\mathbf{g}}_{m,k}\hat{\mathbf{g}}_{m,k}^{\mathrm{H}}\}\\&=\frac{\sigma^4v_{m,k}\overline{\mathbf{g}}_{m,k}\overline{\mathbf{g}}_{m,k}^{\mathrm{H}}}{(\tau_{\mathrm{ul}} p_{\mathrm{ul}}u_{m,k}+\sigma^2)(\tau_{\mathrm{ul}} p_{\mathrm{ul}}u_{m,k}+\sigma^2+N\tau_{\mathrm{ul}}p_{\mathrm{ul}}v_{m,k})}\\&+\frac{\sigma^2u_{m,k}}{\tau_{\mathrm{ul}} p_{\mathrm{ul}}u_{m,k}+\sigma^2}\mathbf{I}_N.
		\end{aligned}
	\end{equation}
	\indent Based on the MSE matrix, the normalized MSE (NMSE) of the estimated channel can be expressed as
	\begin{equation}
		\begin{aligned}
			&\mathrm{NMSE}_{\mathbf{g}_{m,k}}=\frac{\mathrm{Tr}\{\mathbf{MSE}_{\mathbf{g}_{m,k}}\}}{\mathrm{Tr}\{\mathrm{Cov}\{\mathbf{g}_{m,k},\mathbf{g}_{m,k}\}\}}\\&=\frac{\tau_{\mathrm{ul}}p_{\mathrm{ul}}\sigma^2u_{m,k}^2+\sigma^4v_{m,k}+\sigma^4u_{m,k}+N\tau_{\mathrm{ul}}p_{\mathrm{ul}}\sigma^2v_{m,k}u_{m,k}}{u_{m,k}(\tau_{\mathrm{ul}} p_{\mathrm{ul}}u_{m,k}+\sigma^2)(\tau_{\mathrm{ul}} p_{\mathrm{ul}}u_{m,k}+\sigma^2+N\tau_{\mathrm{ul}}p_{\mathrm{ul}}v_{m,k})}.
		\end{aligned}
	\end{equation}
	\subsection{Estimate the Channel Between the APs}
	\indent  In the DL pilot training phase, all the DL APs send DL pilot to the UL APs, the pilot sent by the $m$-th DL AP is denoted as $\boldsymbol{\Phi}_m \in \mathbb{C}^{\tau_{\mathrm{dl}} \times N}$ which satisfies $\left\| \boldsymbol{\Phi}_m \right\|_F^2=1$ and $\tau_{\mathrm{dl}}\geq M_{\mathrm{dl}}$. The pilot signal received at the $l$-th UL AP can be expressed as
	\begin{equation}
		\mathbf{Y}_{l}=\displaystyle\sum_{m=1}^{M_\mathrm{dl}}\sqrt{\tau_{\mathrm{dl}} p_{\mathrm{dl}}}\mathbf{H}_{m,l}\boldsymbol{\Phi}_m^{\mathrm{H}}+\mathbf{N}_l,
	\end{equation}
	where $\tau_{\mathrm{dl}} p_{\mathrm{dl}}$ denotes the transmit DL pilot power.\\
	\indent Multiplying \eqref{1} by $\boldsymbol{\Phi}_m$ and utilizing the orthogonality of the pilot signals, the observation vector of the $m$-th AP at the $l$-th UE can be expressed as follows
	\begin{equation}
		\mathbf{Y}_{m,l}=\sqrt{\tau_{\mathrm{dl}} p_{\mathrm{dl}}}\mathbf{H}_{m,l}+\mathbf{N}_l\boldsymbol{\Phi}_m.
	\end{equation}
	\indent The MMSE estimation of $\mathbf{H}_{m,l}$ can be expressed as:
	\begin{equation}
		\hat{\mathbf{H}}_{m,l}=\mathbf{R}_{\mathbf{H}_{m,l}\mathbf{Y}_{m,l}}\mathbf{R}_{\mathbf{Y}_{m,l}\mathbf{Y}_{m,l}}^{-1}\mathbf{Y}_{m,l},
	\end{equation}
	where
	\begin{equation}
		\begin{aligned}
			\mathbf{R}_{\mathbf{Y}_{m,l}\mathbf{Y}_{m,l}}=\mathbb{E}\{\mathbf{Y}_{m,l}\mathbf{Y}_{m,l}^{\mathrm{H}}\}=(\tau_{\mathrm{dl}} p_{\mathrm{dl}}\gamma_{m,l}+\sigma^2)\mathbf{I}_{N},
		\end{aligned}
	\end{equation}
	\begin{equation}
		\begin{aligned}
			\mathbf{R}_{\mathbf{H}_{m,l}\mathbf{Y}_{m,l}}=\mathbb{E}\{\mathbf{H}_{m,l}\mathbf{Y}_{m,l}^{\mathrm{H}}\}=\sqrt{\tau_{\mathrm{dl}} p_{\mathrm{dl}}}\gamma_{m,l}\mathbf{I}_{N}.
		\end{aligned}
	\end{equation}
	\indent Consequently, we have
	\begin{equation}
		\hat{\mathbf{H}}_{m,l}=\frac{\sqrt{\tau_{\mathrm{dl}} p_{\mathrm{dl}}}\gamma_{m,l}}{\tau_{\mathrm{dl}} p_{\mathrm{dl}}\gamma_{m,l}+\sigma^2}\mathbf{Y}_{m,l}\sim \mathcal{CN}(\mathbf{0},\frac{\tau_{\mathrm{dl}} p_{\mathrm{dl}}\gamma_{m,l}^2}{\tau_{\mathrm{dl}} p_{\mathrm{dl}}\gamma_{m,l}+\sigma^2}\mathbf{I}_N).
	\end{equation}
	\indent The estimation error of $\mathbf{H}_{m,l}$ is denoted as $\mathbf{e}_{m,l}^{\mathrm{AP}}=\mathbf{H}_{m,l}-	\hat{\mathbf{H}}_{m,l}$, then the MSE matrix of the estimated channel is calculated as follows
	\begin{equation}
		\begin{aligned}
			&\mathbf{MSE}_{\mathbf{H}_{m,l}}=\mathbb{E}\{\mathbf{e}_{m,l}^{\mathrm{AP}}(\mathbf{e}_{m,l}^{\mathrm{AP}})^{\mathrm{H}}\}\\&=\mathbb{E}\{\mathbf{H}_{m,l}\mathbf{H}_{m,l}^{\mathrm{H}}\}-\mathbb{E}\{\hat{\mathbf{H}}_{m,l}\hat{\mathbf{H}}_{m,l}^{\mathrm{H}}\}=\frac{\sigma^2\gamma_{m,l}}{\tau_{\mathrm{dl}} p_{\mathrm{dl}}\gamma_{m,l}+\sigma^2}\mathbf{I}_N.
		\end{aligned}
	\end{equation}
	\indent Based on the MSE matrix, the NMSE of the estimated channel can be expressed as
	\begin{equation}
		\mathrm{NMSE}_{\mathbf{H}_{m,l}}=\frac{\mathrm{Tr}\{\mathbf{MSE}_{\mathbf{H}_{m,l}}\}}{\mathrm{Tr}\{\mathrm{Cov}\{\mathbf{H}_{m,l},\mathbf{H}_{m,l}\}\}}=\frac{\sigma^2}{\tau_{\mathrm{dl}} p_{\mathrm{dl}}\gamma_{m,l}+\sigma^2}.
	\end{equation}
	\section{Analysis of communication and sensing performance}\label{Analysis}
	\subsection{Transmit Signal Model}
	\indent All the APs are able to switch between the UL and DL modes. The mode selection strategy is decided based on the long-term statistical CSI as will be discussed in Section \ref{AP Mode Selection Algorithm}. The binary variables to indicate the mode assignment for the $m$-th AP are defined as
	\begin{equation}
		a_m=\left\{
		\begin{array}{cl}
			1, & \text{if the $m$-th AP operates in the DL mode,}\\
			0, & \text{otherwise.}\\
		\end{array} \right.
	\end{equation}
	\begin{equation}
		b_m=\left\{
		\begin{array}{cl}
			1, & \text{if the $m$-th AP operates in the UL mode,}\\
			0, & \text{otherwise.}\\
		\end{array} \right.
	\end{equation}
	which satisfy $a_m+b_m=1$ to guarantee that the $m$-th AP only operates in either the DL or UL mode.\\
	\indent The DL signal transmitted by the $m$-th AP can be expressed as
	\begin{equation}
		\mathbf{x}_m=a_m\left(\displaystyle\sum_{k=1}^{K_{\mathrm{dl}}}\sqrt{p_{\mathrm{dl},m,k}}\mathbf{w}_{m,k}s_{\mathrm{dl},k}+\displaystyle\sum_{t=1}^{K_{\mathrm{t}}}\sqrt{p_{\mathrm{s},m,t}}\mathbf{f}_{m,t}s_{\mathrm{s},t}\right),
	\end{equation}
	where $p_{\mathrm{dl},m,k}$ denotes the transmit power of the $m$-th AP assigned to the $k$-th DL UE, $p_{\mathrm{s},m,t}$ denotes the transmit power of the $m$-th AP assigned to the $l$-th target, $s_{\mathrm{dl},k}$ denotes the communication symbol transmitted to the $k$-th DL UE, $s_{\mathrm{s},t}$ denotes the sensing signal transmitted by the $m$-th AP to detect the $l$-th target, and $\mathbf{w}_{m,k}$ denotes the communication transmit beamforming, which is designed according to the estimated channel based on the low-complexity MRT beamforming as follows
	\begin{equation}
		\mathbf{w}_{m,k}=\frac{\hat{\mathbf{g}}_{m,k}}{\left\| \hat{\mathbf{g}}_{m,k} \right\|_2}.
	\end{equation}
	\indent $\mathbf{f}_{m,t}=\mathbf{a}(\hat{\theta}^{\mathrm{s}}_{m,t},\hat{\phi}^{\mathrm{s}}_{m,t})$ denotes the transmit beamforming for sensing, where
	%\begin{equation}
	%	\mathbf{f}_{m,t}=\mathbf{a}(\hat{\theta}_{m,t},\hat{\phi}_{m,t})
	%\end{equation}
	$\hat{\theta}_{m,t}^{\mathrm{s}}={\theta}_{m,t}^{\mathrm{s}}+\Delta_{\theta}$ and $\hat{\phi}_{m,t}^{\mathrm{s}}={\phi}_{m,t}^{\mathrm{s}}+\Delta_{\phi}$.  ${\theta}_{m,t}^{\mathrm{s}}$ and $\hat{\phi}_{m,t}^{\mathrm{s}}$ denote the azimuth angle and the elevation angle between the $m$-th AP and the $t$-th target. $\Delta_{\theta}$ and $\Delta_{\phi}$ denote the uncertainty of beam alignment. 
	\subsection{Downlink Communication Performance}
	\indent The signal received by the $k$-th DL UE can be expressed as
	\begin{equation}\label{dl}
		\begin{aligned}
		 &y_{\mathrm{dl},k}=\displaystyle\sum_{m=1}^{M}\mathbf{g}_{m,k}^{\mathrm{H}}\mathbf{x}_m+\displaystyle\sum_{j\in\mathcal{K}_{\mathrm{ul}}}\sqrt{p_{\mathrm{ul}}}h_{k,j}s_j+n_k\\&=\underbrace{\displaystyle\sum_{m=1}^{M}a_m\mathbf{g}_{m,k}^{\mathrm{H}}\sqrt{p_{\mathrm{dl},m,k}}\mathbf{w}_{m,k}s_{\mathrm{dl},k}}_{\text{desired signal}}+\underbrace{\displaystyle\sum_{j\in\mathcal{K}_{\mathrm{ul}}}\sqrt{p_{\mathrm{ul}}}h_{k,j}s_j}_{\text{UL UEs interference}}\\&+\underbrace{\displaystyle\sum_{m=1}^{M}a_m\mathbf{g}_{m,k}^{\mathrm{H}}\displaystyle\sum_{i=1,i\neq k}^{K_{\mathrm{dl}}}\sqrt{p_{\mathrm{dl},m,i}}\mathbf{w}_{m,i}s_{\mathrm{dl},i}}_{\text{\text{multi-user interference}}}\\&+\underbrace{\displaystyle\sum_{m=1}^{M}a_m\mathbf{g}_{m,k}^{\mathrm{H}}\displaystyle\sum_{t=1}^{K_{\mathrm{t}}}\sqrt{p_{\mathrm{s},m,t}}\mathbf{f}_{m,t}s_{\mathrm{s},t}}_{\text{sensing symbol interference}}+n_k,
		\end{aligned}
	\end{equation}
	where $n_k\sim \mathcal{CN}(0,\sigma^2)$ denotes the AWGN at the $k$-th DL UE, and $p_{\mathrm{ul}}$ denotes the transmit power of each UL UE. To evaluate the performance of DL communication, closed-form expression of the ergodic rate of the $k$-th DL UE is derived based on the statistical CSI as follows
	\begin{equation}\label{dl_rate}
		R_{\mathrm{dl},k}=\overline{\tau}\log_2\left(1+\frac{\left(\displaystyle\sum_{m=1}^{M}c_{\mathrm{dl},m,k}^{(1)}a_m\right)^2}{\displaystyle\sum_{m=1}^{M}c_{\mathrm{dl},m,k}^{(2)}a_m^2+c_{\mathrm{dl},k}^{(3)}}\right),
	\end{equation}
	\begin{figure*}[t]
		\begin{equation}\label{DL1}
			c_{\mathrm{dl},m,k}^{(1)}=\zeta_{m,k}\sqrt{p_{\mathrm{dl},m,k}\tau_{\mathrm{ul}}p_{\mathrm{ul}}}\left(v_{m,k}\overline{\mathbf{g}}_{m,k}^{\mathrm{H}}\mathbf{A}_{m,k}\overline{\mathbf{g}}_{m,k}+u_{m,k}\mathrm{Tr}(\mathbf{A}_{m,k})\right),\quad c_{\mathrm{dl},k}^{(3)}=\displaystyle\sum_{j\in\mathcal{K}_{\mathrm{ul}}}p_{\mathrm{ul}}\alpha_{k,j}+\sigma^2,
		\end{equation}
		\begin{equation}\label{DL2}
			\begin{aligned}
			&c_{\mathrm{dl},m,k}^{(2)}=\zeta_{m,k}^2p_{\mathrm{dl},m,k}\tau_{\mathrm{ul}}p_{\mathrm{ul}}\Big(v_{m,k}^2\overline{\mathbf{g}}_{m,k}^{\mathrm{H}}\mathbf{A}_{m,k}\overline{\mathbf{g}}_{m,k}\overline{\mathbf{g}}_{m,k}^{\mathrm{H}}\mathbf{A}_{m,k}^{\mathrm{H}}\overline{\mathbf{g}}_{m,k}+2u_{m,k}v_{m,k}\mathrm{Tr}\{\mathbf{A}_{m,k}\overline{\mathbf{g}}_{m,k}\overline{\mathbf{g}}_{m,k}^{\mathrm{H}}\mathbf{A}_{m,k}^{\mathrm{H}}\}\\&+2u_{m,k}v_{m,k}\overline{\mathbf{g}}_{m,k}^{\mathrm{H}}\mathbf{A}_{m,k}\overline{\mathbf{g}}_{m,k}\mathrm{Tr}\{\mathbf{A}_{m,k}\}+u_{m,k}^2(\mathrm{Tr}\{\mathbf{A}_{m,k}^2\}+|\mathrm{Tr}\{\mathbf{A}_{m,k}\}|^2)-\left(v_{m,k}\overline{\mathbf{g}}_{m,k}^{\mathrm{H}}\mathbf{A}_{m,k}\overline{\mathbf{g}}_{m,k}+u_{m,k}\mathrm{Tr}(\mathbf{A}_{m,k})\right)^2\Big)\\&+\displaystyle\sum_{i=1,i\neq k}^{K_{\mathrm{dl}}}\zeta_{m,i}^2p_{\mathrm{dl},m,i}\left(v_{m,k}\overline{\mathbf{g}}_{m,k}^{\mathrm{H}}\mathbf{B}_{m,i}\overline{\mathbf{g}}_{m,k}+u_{m,k}\mathrm{Tr}(\mathbf{B}_{m,i})\right)+v_{m,k}\overline{\mathbf{g}}_{m,k}^{\mathrm{H}}\mathbf{R}_m\overline{\mathbf{g}}_{m,k}+u_{m,k}\mathrm{Tr}(\mathbf{R}_m).
			\end{aligned}
		\end{equation}
		\centering
		\vspace*{0pt}
		\hrulefill
		\vspace*{0pt} 
	\end{figure*}
where $\overline{\tau}=\frac{\tau_c-\tau_{\mathrm{ul}}-\tau_{\mathrm{dl}}}{\tau_c}$.\\
\indent By defining $\zeta_{m,k}^2\triangleq1/\mathbb{E}\{\left\| \hat{\mathbf{g}}_{m,k} \right\|_2^2\}$,  $\mathbf{B}_{m,i}\triangleq\mathbf{A}_{m,i}\left(\tau_{\mathrm{ul}}p_{\mathrm{ul}}v_{m,i}\overline{\mathbf{g}}_{m,i}\overline{\mathbf{g}}_{m,i}^{\mathrm{H}}+\left(\tau_{\mathrm{ul}}p_{\mathrm{ul}}u_{m,i}+\sigma^2\right)\mathbf{I}_N\right)\mathbf{A}_{m,i}^{\mathrm{H}}$, and $\mathbf{R}_{m}\triangleq\displaystyle\sum_{t=1}^{K_{\mathrm{t}}}p_{\mathrm{s},m,t}\mathbf{f}_{m,t}\mathbf{f}_{m,t}^{\mathrm{H}}$. $c_{\mathrm{dl},m,k}^{(1)}$, $c_{\mathrm{dl},m,k}^{(2)}$, and $c_{\mathrm{dl},m,k}^{(3)}$ are constants that are only relevant to the large-scale information, whose expressions are given by \eqref{DL1} and \eqref{DL2}. The detailed derivations are provided in Appendix \ref{B}.
	\subsection{Uplink Communication Performance}
	\indent For the UL data transmission, the received signal at the $m$-th AP can be expressed as
	\begin{equation}
		\begin{aligned}
			\mathbf{y}_{\mathrm{ul},m}&=b_m\left(\displaystyle\sum_{j\in\mathcal{K}_{\mathrm{ul}}}\sqrt{p_{\mathrm{ul}}}\mathbf{g}_{m,j}s_j+\displaystyle\sum_{l=1}^{M}\mathbf{H}_{m,l}\mathbf{x}_l+\mathbf{y}_{\mathrm{e},m}+\mathbf{n}_m\right),
		\end{aligned}
	\end{equation}
	where $\mathbf{n}_m\sim\mathcal{CN}(\mathbf{0},\sigma^2\mathbf{I}_N)$ denotes the AWGN at the $m$-th AP, and $	\mathbf{y}_{\mathrm{e},m}=\displaystyle\sum_{t=1}^{K_{\mathrm{t}}}\displaystyle\sum_{l=1}^{M}\mathbf{G}_{m,t,l}\mathbf{x}_l$ denotes the summation of the echo signal from the $K_{\mathrm{t}}$ targets received by the $m$-th AP, with $	\mathbf{G}_{m,t,l}=\sqrt{\rho_{m,t,l}}\mathbf{a}(\theta_{m,t},\phi_{m,t})\mathbf{a}^{\mathrm{H}}(\theta_{l,t},\phi_{l,t})$ being the target response matrix and $\rho_{m,t,l}$ being the radar pathloss.\\
	\indent By using the MRC detector, the signal from the $i$-th UL UE at the $m$-th AP can be expressed as
	\begin{equation}
		\begin{aligned}
			y_{\mathrm{ul},m,i}&=b_m\sqrt{p_{\mathrm{ul}}}\hat{\mathbf{g}}_{m,i}^{\mathrm{H}}\mathbf{g}_{m,i}s_i+\displaystyle\sum_{j\in\mathcal{K}\backslash\{i\}}b_m\sqrt{p_{\mathrm{ul}}}\hat{\mathbf{g}}_{m,i}^{\mathrm{H}}\mathbf{g}_{m,j}s_j\\&+\displaystyle\sum_{l=1}^{M}b_m\hat{\mathbf{g}}_{m,i}^{\mathrm{H}}\mathbf{H}_{m,l}\mathbf{x}_l+b_m\hat{\mathbf{g}}_{m,i}^{\mathrm{H}}\mathbf{y}_{\mathrm{e},m}+b_m\hat{\mathbf{g}}_{m,i}^{\mathrm{H}}\mathbf{n}_{m}.
		\end{aligned}
	\end{equation}
	\indent  Because the signals transmitted by the DL APs can be reconstructed, the interference between the APs can be partially mitigated. Consequently, the signal after interference mitigation can be expressed as 
	\begin{equation}\label{ul}
		\begin{aligned}
			&y_{\mathrm{ul},m,i}=\underbrace{b_m\sqrt{p_{\mathrm{ul}}}\hat{\mathbf{g}}_{m,i}^{\mathrm{H}}\mathbf{g}_{m,i}s_i}_{\text{desired signal}}+\underbrace{\displaystyle\sum_{j\in\mathcal{K}_{\mathrm{ul}}\backslash\{i\}}b_m\sqrt{p_{\mathrm{ul}}}\hat{\mathbf{g}}_{m,i}^{\mathrm{H}}\mathbf{g}_{m,j}s_j}_{\text{multi-user interference}}\\&+\underbrace{\displaystyle\sum_{l=1}^{M}b_m\hat{\mathbf{g}}_{m,i}^{\mathrm{H}}(\mathbf{H}_{m,l}-\hat{\mathbf{H}}_{m,l})\mathbf{x}_l}_{\text{residual interference from the DL APs}}+\underbrace{b_m\hat{\mathbf{g}}_{m,i}^{\mathrm{H}}\mathbf{y}_{\mathrm{e},m}}_{\text{echo interference}}+b_m\hat{\mathbf{g}}_{m,i}^{\mathrm{H}}\mathbf{n}_{m}.
		\end{aligned}
	\end{equation}
	\indent Closed-form expression of the ergodic rate of the $i$-th UL UE is derived based on the statistical CSI as follows
	\begin{equation}\label{ul_rate}
		\begin{aligned}	&R_{\mathrm{ul},i}=\overline{\tau}\log_2\left(1+\frac{\left(\displaystyle\sum_{m=1}^{M}c_{\mathrm{ul},m,i}^{(1)}b_m\right)^2}{\displaystyle\sum_{m=1}^{M}c_{\mathrm{ul},m,i}^{(2)}b_m^2+\displaystyle\sum_{m=1}^{M}\displaystyle\sum_{l=1}^{M}c_{\mathrm{ul},m,l,i}^{(3)}a_l^2b_m^2}\right),
		\end{aligned}
	\end{equation}
	\begin{figure*}[b]
		\centering
		\vspace*{0pt}
		\hrulefill
		\vspace*{0pt} 
		\begin{equation}\label{UL1}
			c_{\mathrm{ul},m,i}^{(1)}=\sqrt{\tau_{\mathrm{ul}}p_{\mathrm{ul}}^2}\left(v_{m,i}\overline{\mathbf{g}}_{m,i}^{\mathrm{H}}\mathbf{A}_{m,i}\overline{\mathbf{g}}_{m,i}+u_{m,i}\mathrm{Tr}(\mathbf{A}_{m,i})\right),
		\end{equation}
		\begin{equation}\label{UL2}
			\begin{aligned}
				c_{\mathrm{ul},m,i}^{(2)}&=\sigma^2\zeta_{m,i}^2+\tau_{\mathrm{ul}}p_{\mathrm{ul}}^2\Big (v_{m,i}^2\overline{\mathbf{g}}_{m,i}^{\mathrm{H}}\mathbf{A}_{m,i}\overline{\mathbf{g}}_{m,i}\overline{\mathbf{g}}_{m,i}^{\mathrm{H}}\mathbf{A}_{m,i}^{\mathrm{H}}\overline{\mathbf{g}}_{m,i}+2u_{m,i}v_{m,i}\mathrm{Tr}\{\mathbf{A}_{m,i}\overline{\mathbf{g}}_{m,i}\overline{\mathbf{g}}_{m,i}^{\mathrm{H}}\mathbf{A}_{m,i}^{\mathrm{H}}\}\\&+2u_{m,i}v_{m,i}\overline{\mathbf{g}}_{m,i}^{\mathrm{H}}\mathbf{A}_{m,i}\overline{\mathbf{g}}_{m,i}\mathrm{Tr}\{\mathbf{A}_{m,i}\}+u_{m,i}^2(\mathrm{Tr}\{\mathbf{A}_{m,i}^2\}+|\mathrm{Tr}\{\mathbf{A}_{m,i}\}|^2)\\&-\left(v_{m,i}\overline{\mathbf{g}}_{m,i}^{\mathrm{H}}\mathbf{A}_{m,i}\overline{\mathbf{g}}_{m,i}+u_{m,i}\mathrm{Tr}(\mathbf{A}_{m,i})\right)^2\Big)+\displaystyle\sum_{j\in\mathcal{K}\backslash\{i\}}p_{\mathrm{ul}}\left(v_{m,j}\overline{\mathbf{g}}_{m,j}^{\mathrm{H}}\mathbf{B}_{m,i}\overline{\mathbf{g}}_{m,j}+u_{m,j}\mathrm{Tr}(\mathbf{B}_{m,i})\right),
			\end{aligned}
		\end{equation}
		\begin{equation}\label{UL3}
			\begin{aligned}
				c_{\mathrm{ul},m,l,i}^{(3)}&=\frac{\sigma^2\gamma_{m,l}\zeta_{m,i}^2}{\tau_{\mathrm{dl}} p_{\mathrm{dl}}\gamma_{m,l}+\sigma^2}\left(\displaystyle\sum_{k=1}^{K_{\mathrm{dl}}}\zeta_{l,k}^2p_{\mathrm{dl},l,k}\mathrm{Tr}(\mathbf{B}_{l,k})+\mathrm{Tr}(\mathbf{R}_{l})\right)\\&+\tau_{\mathrm{ul}} p_{\mathrm{ul}}\Big(v_{m,i}\overline{\mathbf{g}}_{m,i}^{\mathrm{H}}\mathbf{A}_{m,i}^{\mathrm{H}}\mathbf{X}_{m,l}\mathbf{A}_{m,i}\overline{\mathbf{g}}_{m,i}+u_{m,i}\mathrm{Tr}(\mathbf{A}_{m,i}^{\mathrm{H}}\mathbf{X}_{m,l}\mathbf{A}_{m,i})\Big)+\sigma^2\mathrm{Tr}(\mathbf{A}_{m,i}^{\mathrm{H}}\mathbf{X}_{m,l}\mathbf{A}_{m,i}).
			\end{aligned}
		\end{equation}
	\end{figure*}
\noindent where $c_{\mathrm{ul},m,i}^{(1)}$, $c_{\mathrm{ul},m,i}^{(2)}$, and $c_{\mathrm{ul},m,l,i}^{(3)}$ are constants that are only relevant to the large-scale information, whose expressions are given by \eqref{UL1}, \eqref{UL2}, and \eqref{UL3}. The expression of $\mathbf{X}_{m,l}$ in \eqref{UL3} is given by $\mathbf{X}_{m,l}=\displaystyle\sum_{t=1}^{K_{\mathrm{t}}}\mathbf{G}_{m,t,l}(\displaystyle\sum_{k=1}^{K_{\mathrm{dl}}}\zeta_{l,k}^2p_{\mathrm{dl},l,k}\mathbf{B}_{l,k}+\mathbf{R}_{l})\mathbf{G}^{\mathrm{H}}_{m,t,l}$. The detailed derivations are provided in Appendix \ref{C}.
	\subsection{Sensing Performance}
\indent For target sensing, the received signal at the $m$-th AP can be expressed as
	\begin{equation}
		\begin{aligned}
			\mathbf{y}_{\mathrm{s},m}&=b_m\displaystyle\sum_{t=1}^{K_{\mathrm{t}}}\displaystyle\sum_{l=1}^{M}\mathbf{G}_{m,t,l}\mathbf{x}_l+b_m\displaystyle\sum_{j\in\mathcal{K}_{\mathrm{ul}}}\sqrt{p_{\mathrm{ul}}}\mathbf{g}_{m,j}s_j\\&+b_m\displaystyle\sum_{l=1}^{M}\mathbf{H}_{m,l}\mathbf{x}_l+b_m\mathbf{n}_m.
		\end{aligned}
	\end{equation}
\indent For performing target sensing, the significant radar passloss leads to weak echo signal. To deal with the strong interference from the UL UEs and the DL APs, we use the estimated channels to suppress the interference. After interference cancellation, the total interference together with noise can be expressed as
	\begin{equation}
		\begin{aligned}
			\mathbf{n}_{\mathrm{s},m}=\displaystyle\sum_{j\in\mathcal{K}_{\mathrm{ul}}}&\sqrt{p_{\mathrm{ul}}}(\mathbf{g}_{m,j}-\hat{\mathbf{g}}_{m,j})s_j\\&+\displaystyle\sum_{l=1}^{M}(\mathbf{H}_{m,l}-\hat{\mathbf{H}}_{m,l})\mathbf{x}_l+\mathbf{n}_m.
		\end{aligned}
	\end{equation}
	\indent After being processed by the receive filter $\mathbf{u}_m=[\mathbf{u}_{m,1},\dots,\mathbf{u}_{m,K_{\mathrm{t}}}]$ for $\mathbf{u}_{m,t}=\mathbf{a}(\hat{\theta}_{m,t},\hat{\phi}_{m,t})$, the radar output signal at the $m$-th AP for sensing the $p$-th target is obtained as follows
	\begin{equation}
		\begin{aligned}\label{sen}
			y_{\mathrm{s},m,p}&=b_m\mathbf{u}_{m,p}^{\mathrm{H}}\displaystyle\sum_{l=1}^{M}a_l\mathbf{G}_{m,p,l}\mathbf{x}_l\\&+b_m\mathbf{u}_{m,p}^{\mathrm{H}}\displaystyle\sum_{t=1,t\neq p}^{K_{\mathrm{t}}}\displaystyle\sum_{l=1}^{M}a_l\mathbf{G}_{m,t,l}\mathbf{x}_l+b_m\mathbf{u}_{m,p}^{\mathrm{H}}\mathbf{n}_{\mathrm{s},m}.
		\end{aligned}
	\end{equation}
	\indent Thus, the radar output signal gathered at the CPU for sensing the $p$-th target can be expressed as
	\begin{equation}
		\begin{aligned}
			\mathbf{y}_{\mathrm{s},p}=[y_{\mathrm{s},1,p},\dots,y_{\mathrm{s},M,p}]^{\mathrm{T}}.
		\end{aligned}
	\end{equation}
	\indent Thus, the sensing SINR of the $p$-th target can be expressed as
	\begin{equation}\label{sinr}
		\mathrm{SINR}_{\mathrm{s},p}=\frac{\displaystyle\sum_{m=1}^{M}\displaystyle\sum_{l=1}^{M}c_{\mathrm{s},m,p,l}^{(1)}a_l^2b_m^2}{\displaystyle\sum_{m=1}^{M}\displaystyle\sum_{l=1}^{M}c_{\mathrm{s},m,p,l}^{(2)}a_l^2b_m^2+\displaystyle\sum_{m=1}^{M}c_{\mathrm{s},m,p}^{(3)}b_m^2},
	\end{equation}
	where $c_{\mathrm{s},m,p,l}^{(1)}$, $c_{\mathrm{s},m,p,l}^{(2)}$, and $c_{\mathrm{s},m,p,l}^{(3)}$ are constants that are only relevant to the large-scale information, whose expressions are given by
	\begin{equation}
		\begin{aligned}
			c_{\mathrm{s},m,p,l}^{(1)}=\mathbf{u}_{m,p}^{\mathrm{H}}\mathbf{G}_{m,p,l}(\displaystyle\sum_{i=1}^{K_{\mathrm{dl}}}\zeta_{l,i}^2p_{\mathrm{dl},l,i}\mathbf{B}_{l,i}+\mathbf{R}_{l})\mathbf{G}_{m,p,l}^{\mathrm{H}}\mathbf{u}_{m,p},
		\end{aligned}
	\end{equation}
	\begin{equation}
		\begin{aligned}
			&c_{\mathrm{s},m,p,l}^{(2)}\\&=\displaystyle\sum_{t=1,t\neq p}^{K_{\mathrm{t}}}\mathbf{u}_{m,p}^{\mathrm{H}}\mathbf{G}_{m,t,l}(\displaystyle\sum_{i=1}^{K_{\mathrm{dl}}}\zeta_{l,i}^2p_{\mathrm{dl},l,i}\mathbf{B}_{l,i}+\mathbf{R}_{l})\mathbf{G}_{m,t,l}^{\mathrm{H}}\mathbf{u}_{m,p}\\&+\frac{N\sigma^2\gamma_{m,l}}{\tau_{\mathrm{dl}} p_{\mathrm{dl}}\gamma_{m,l}+\sigma^2}\left(\displaystyle\sum_{i=1}^{K_{\mathrm{dl}}}\zeta_{l,i}^2p_{\mathrm{dl},l,i}\mathrm{Tr}(\mathbf{B}_{l,i})+\mathrm{Tr}(\mathbf{R}_{l})\right),
		\end{aligned}
	\end{equation}
	\begin{equation}
		\begin{aligned}
			c_{\mathrm{s},m,p,l}^{(3)}=\displaystyle\sum_{j\in\mathcal{K}_{\mathrm{ul}}}p_{\mathrm{ul}}Nu_{m,j}\mathrm{NMSE}_{\mathbf{g}_{m,j}}+N\sigma^2.
		\end{aligned}
	\end{equation}
	\indent The detailed derivations are provided in Appendix \ref{D}.
	\section{AP Mode Selection Algorithm}\label{AP Mode Selection Algorithm}
	\indent According to the analytical results given in \eqref{dl_rate}, \eqref{ul_rate} and \eqref{sinr}, we aim to design the AP mode selection to maximize the performance of the cooperative ISAC network via the long-term statistical CSI.
	\subsection{Problem Formulation}
	\indent We aim to optimize the UL and DL mode assignment $(\bm{a},\bm{b})$ to maximize the minimum sensing SINR of all the targets while satisfying the communication QoS requirements of all the DL and UL UEs. The problem is formulated as follows
	\begin{subequations}\label{P1}
		\begin{align}
			\mathop {\max }\limits_{ {\bm{a}}, {\bm{b}} }
			& \min_{p\in\mathcal{K}_{\mathrm{t}}}\quad \mathrm{SINR}_{\mathrm{s},p}
			\\
			\ \textrm{s.t.} \quad
			&R_{\mathrm{dl},k}\geq \kappa_{\mathrm{dl}},\forall k\in \mathcal{K}_{\mathrm{dl}},\label{P1a}\\
			& R_{\mathrm{ul},i}\geq \kappa_{\mathrm{ul}},\forall i\in \mathcal{K}_{\mathrm{ul}}, \label{P1b}\\
			& a_m+b_m=1, m=1,\dots,M, \label{P1c}\\
			& a_m,b_m\in\{0,1\}, m=1,\dots,M. \label{P1d}
		\end{align}
	\end{subequations}
	\indent Problem \eqref{P1} is an NP-hard integer optimization problem due to the binary variables involved and non-convexity of the objective function and the communication QoS constraints. Moreover, there exists a strong coupling between $\bm{a}$ and $\bm{b}$, which makes Problem \eqref{P1} even more complicated. In what follows, we transform Problem \eqref{P1} into a more tractable form by exploiting the special property of binary variables, and use SCA techniques to solve the problem efficiently.
	\subsection{Solution of the Optimization Problem}
	\indent By introducing slack variable $t=\min_{p\in\mathcal{K}_{\mathrm{t}}}\mathrm{SINR}_{\mathrm{s},p}$, Problem \eqref{P1} can be transformed into a more intractable form as follows
	\begin{subequations}\label{P2}
		\begin{align}
			\mathop {\min }\limits_{ {\bm{a}}, {\bm{b}}, t }
			& \qquad -t
			\\
			\ \textrm{s.t.} \quad
			&\mathrm{SINR}_{\mathrm{s},p} \geq t,\forall p\in \mathcal{K}_{\mathrm{t}},\label{P2a}\\
			&R_{\mathrm{dl},k}\geq \kappa_{\mathrm{dl}},\forall k\in \mathcal{K}_{\mathrm{dl}},\label{P2b}\\
			& R_{\mathrm{ul},i}\geq \kappa_{\mathrm{ul}},\forall i\in \mathcal{K}_{\mathrm{ul}}, \label{P2c}\\
			& a_m+b_m=1, m=1,\dots,M, \label{P2d}\\
			& a_m,b_m\in\{0,1\}, m=1,\dots,M. \label{P2e}
		\end{align}
	\end{subequations}
	\indent Because $a_m,b_m\in\{0,1\}$, resulting in $a_m^2=a_m$ and $b_m^2=b_m$, so we first replace $a_m^2,b_m^2$ with $a_m,b_m$ in Problem \eqref{P2}. Furthermore, by introducing a collection of variables ${\bm{\mu}}=[\mu_1,\dots,\mu_{K_{\mathrm{t}}}]^{\mathrm{T}}\in \mathbb{R}^{K_{\mathrm{t}}\times 1}$, constraint \eqref{P2a} can be converted into 
	\begin{equation}\label{C1}
		\begin{aligned}
			\mu_p^2\Big(\displaystyle\sum_{m=1}^{M}\displaystyle\sum_{l=1}^{M}c_{\mathrm{s},m,p,l}^{(2)}&a_lb_m+\displaystyle\sum_{m=1}^{M}c_{\mathrm{s},m,p}^{(3)}b_m\Big)\\&-2\mu_p\sqrt{\displaystyle\sum_{m=1}^{M}\displaystyle\sum_{l=1}^{M}c_{\mathrm{s},m,p,l}^{(1)}a_lb_m}+t \leq 0, \forall p.
		\end{aligned}
	\end{equation}
\indent In each iteration, when ${\bm{a}}$, ${\bm{b}}$, and $t$ are fixed, the optimal value of $\mu_p$ is given by \cite{shen2018fractional}
	\begin{equation} \label{80}
		\mu_p^{*}=\frac{\sqrt{\displaystyle\sum_{m=1}^{M}\displaystyle\sum_{l=1}^{M}c_{\mathrm{s},m,p,l}^{(1)}a_lb_m}}{\displaystyle\sum_{m=1}^{M}\displaystyle\sum_{l=1}^{M}c_{\mathrm{s},m,p,l}^{(2)}a_lb_m+\displaystyle\sum_{m=1}^{M}c_{\mathrm{s},m,p}^{(3)}b_m},p=1,\dots,K_{\mathrm{t}}.
	\end{equation}
	\indent Because ${\bm{a}}$ and ${\bm{b}}$ are strongly coupled binary variables, constraint \eqref{C1} is not convex. To tackle the non-convexity, we propose to use the McCormick envelope \cite{scott2011generalized} to approximate the problem. Specifically, a collection of variables  ${\bm{\chi}}=[\chi_{1,1},\chi_{1,2},\dots,\chi_{M,M-1},\chi_{M,M}]^{\mathrm{T}}\in \mathbb{R}^{M^2 \times 1}$ are introduced, whose elements satisfy\\
	\begin{equation}
		\left\{
		\begin{array}{cl}
			\chi_{m,l} \leq a_{l}\\
			\chi_{m,l} \leq b_{m}\\
			\chi_{m,l} \geq a_{l}+b_{m}-1\\
			\chi_{m,l} \in \{0,1\}
		\end{array} \right.\quad m,l=1,\dots,M.
	\end{equation}
\indent	Thus \eqref{C1} can be transformed into a convex one as follows
	\begin{equation}
		\begin{aligned}\label{C2}
			\mu_p^2\Big(\displaystyle\sum_{m=1}^{M}\displaystyle\sum_{l=1}^{M}&c_{\mathrm{s},m,p,l}^{(2)}\chi_{m,l}+\displaystyle\sum_{m=1}^{M}c_{\mathrm{s},m,p}^{(3)}b_m\Big)\\&-2\mu_p\sqrt{\displaystyle\sum_{m=1}^{M}\displaystyle\sum_{l=1}^{M}c_{\mathrm{s},m,p,l}^{(1)}\chi_{m,l}}+t \leq 0, \forall p.
		\end{aligned}
	\end{equation}
	\indent By defining $\eta_{\mathrm{dl}}\triangleq2^{\frac{\kappa_{\mathrm{dl}}}{\overline{\tau}}-1}$, constraint \eqref{P2c} can be rewritten as follows
	\begin{equation}\label{C3}
		\begin{aligned}
			\eta_{\mathrm{dl}}\left(\displaystyle\sum_{m=1}^{M}c_{\mathrm{dl},m,k}^{(2)}a_m+c_{\mathrm{dl},k}^{(3)}\right)-\left(\displaystyle\sum_{m=1}^{M}c_{\mathrm{dl},m,k}^{(1)}a_m\right)^2 \leq 0, \forall k.
		\end{aligned}	
	\end{equation}
	\indent We adopt the first-order Taylor expansion to obtain the lower bound of $\left(\displaystyle\sum_{m=1}^{M}c_{\mathrm{dl},m,k}^{(1)}a_m\right)^2$. Thus, constraint \eqref{P2c} can be transformed into a convex one as follows
	\begin{equation}
		\begin{aligned}
			&\eta_{\mathrm{dl}}\left(\displaystyle\sum_{m=1}^{M}c_{\mathrm{dl},m,k}^{(2)}a_m+c_{\mathrm{dl},k}^{(3)}\right)-\left(\displaystyle\sum_{m=1}^{M}c_{\mathrm{dl},m,k}^{(1)}a_m^{(n)}\right)^2\\&-\displaystyle\sum_{j=1}^{M}2c_{\mathrm{dl},j,k}^{(1)}\left(\displaystyle\sum_{m=1}^{M}c_{\mathrm{dl},m,k}^{(1)}a_m^{(n)}\right)\left(a_j-a_j^{(n)}\right) \leq 0, \forall k.
		\end{aligned}
	\end{equation}
	\indent By defining $\eta_{\mathrm{ul}}\triangleq2^{\frac{\kappa_{\mathrm{ul}}}{\overline{\tau}}}-1$, constraint \eqref{P2d} can be rewrote as follows
	\begin{equation}
		\begin{aligned}
			\eta_{\mathrm{ul}}\Big(\displaystyle\sum_{m=1}^{M}c_{\mathrm{ul},m,i}^{(2)}b_m+\displaystyle\sum_{m=1}^{M}\displaystyle\sum_{l=1}^{M}&c_{\mathrm{ul},m,l,i}^{(3)}\chi_{m,l}\Big)\\&-\left(\displaystyle\sum_{m=1}^{M}c_{\mathrm{ul},m,i}^{(1)}b_m\right)^2 \leq 0, \forall i.
		\end{aligned}
	\end{equation}
	\indent Similarly, we adopt the first-order Taylor expansion to obtain the lower bound of $\left(\displaystyle\sum_{m=1}^{M}c_{\mathrm{ul},m,i}^{(1)}b_m\right)^2$. Thus, constraint \eqref{P2d} can be transformed into a convex one as follows
	\begin{equation}
		\begin{aligned}\label{C4}
			\eta_{\mathrm{ul}}\Big(\displaystyle\sum_{m=1}^{M}c_{\mathrm{ul},m,i}^{(2)}&b_m+\displaystyle\sum_{m=1}^{M}\displaystyle\sum_{l=1}^{M}c_{\mathrm{ul},m,l,i}^{(3)}\chi_{m,l}\Big)\\&-\displaystyle\sum_{j=1}^{M}2c_{\mathrm{ul},j,i}^{(1)}\left(\displaystyle\sum_{m=1}^{M}c_{\mathrm{ul},m,i}^{(1)}b_m^{(n)}\right)\left(b_j-b_j^{(n)}\right)\\&-\left(\displaystyle\sum_{m=1}^{M}c_{\mathrm{ul},m,i}^{(1)}b_m^{(n)}\right)^2 \leq 0, \forall i.
		\end{aligned}
	\end{equation}
	\indent When $\bm{\mu}$ is fixed, Problem \eqref{P2} can be transformed into the following one
	\begin{subequations}\label{P3}
		\begin{align}
			\mathop {\min }\limits_{ {\bm{a}}, {\bm{b}}, {\bm{\chi}}, t }
			& \qquad -t
			\\
			\ \textrm{s.t.} \quad
			&\eqref{C2},\eqref{C3},\eqref{C4},\label{P3a}\\
			& a_m+b_m=1, m=1,\dots,M, \label{P3b}\\
			& a_m,b_m\in\{0,1\},\chi_{m,l}\in\{0,1\},m,l=1,\dots,M, \label{P3c}\\
			& \chi_{m,l} \leq a_{l}, \chi_{m,l} \leq b_{m} ,m,l=1,\dots,M,\label{P3d}\\
			& \chi_{m,l} \geq a_{l}+b_{m}-1,m,l=1,\dots,M. \label{P3e}
		\end{align}
	\end{subequations}
	\indent To handle the binary constraint \eqref{P3c}, it is observed that for any real number $x$, we have $x\in\{0,1\}\iff x-x^2=0 \iff (x\in[0,1]\;\&\; x-x^2\leq0)$. Thus, \eqref{P3c} can be replaced by the following equivalent constraints:
	\begin{equation}\label{P3_1_1}
		\displaystyle\sum_{m=1}^{M}(a_m-a_m^2+b_m-b_m^2)+\displaystyle\sum_{m=1}^{M}\displaystyle\sum_{l=1}^{M}(\chi_{m,l}-\chi_{m,l}^2)\leq 0,
	\end{equation}
	\begin{equation}\label{P3_1_2}
		0\leq a_m\leq 1,0\leq b_m\leq 1, 0\leq \chi_{m,l}\leq 1,m,l=1,\dots,M.
	\end{equation}
	\indent Thus, Problem \eqref{P3} can be transformed into 
	\begin{subequations}\label{P3_1}
		\begin{align}
			\mathop {\min }\limits_{ {\bm{a}}, {\bm{b}}, {\bm{\chi}}, t }
			& \qquad -t
			\\
			\ \textrm{s.t.} \quad
			&\eqref{C2},\eqref{C3},\eqref{C4},\eqref{P3b},\eqref{P3d},\eqref{P3e},\eqref{P3_1_1}, \eqref{P3_1_2}.\label{P3_1a}
		\end{align}
	\end{subequations}
	\indent By introducing the Lagrangian multiplier $\lambda$ corresponding to constraint \eqref{P3c} and defining $\varphi({\bm{a}}, {\bm{b}}, {\bm{\chi}})\triangleq\Big(\displaystyle\sum_{m=1}^{M}(a_m-a_m^2+b_m-b_m^2)+\displaystyle\sum_{m=1}^{M}\displaystyle\sum_{l=1}^{M}(\chi_{m,l}-\chi_{m,l}^2)\Big)$, we obtain the Lagrangian of Problem \eqref{P3} as follows
	\begin{subequations}\label{P4}
		\begin{align}
			\mathop {\min }\limits_{ {\bm{a}}, {\bm{b}}, {\bm{\chi}}, t }
			& \quad -t+\lambda\varphi({\bm{a}}, {\bm{b}}, {\bm{\chi}})
			\\
			\ \textrm{s.t.} \quad 
			&\eqref{C2},\eqref{C3},\eqref{C4},\eqref{P3b},\eqref{P3d},\eqref{P3e}, \eqref{P3_1_2}.\label{P4a}
		\end{align}
	\end{subequations}
	\indent It can be proved that the value of $\varphi({\bm{a}}, {\bm{b}}, {\bm{\chi}})$ corresponding to $\lambda$ is decreasing to 0 as $\lambda\to +\infty$, and Problem $\eqref{P3_1}$ has strong duality. Therefore it is equivalent to Problem \eqref{P4} at the optimal solution $\lambda^{*} \geq 0$ of its sup-min problem \cite{vu2018spectral}.\\
	\indent The convex upper bound of $\varphi({\bm{a}}, {\bm{b}}, {\bm{\chi}})$ is given by
	\begin{equation}
		\begin{aligned}
		&\hat{\varphi}({\bm{a}}, {\bm{b}}, {\bm{\chi}})=\displaystyle\sum_{m=1}^{M}(a_m-2a_m^{(n)}a_m+(a_m^{(n)})^2+b_m-2b_m^{(n)}b_m\\&+(b_m^{(n)})^2)+\displaystyle\sum_{m=1}^{M}\displaystyle\sum_{l=1}^{M}(\chi_{m,l}-2\chi_{m,l}^{(n)}\chi_{m,l}+(\chi_{m,l}^{(n)})^2),
		\end{aligned}
	\end{equation}
	where $a_m^{(n)}$, $b_m^{(n)}$ and $\chi_{m,l}^{(n)}$ denote the values at the $n$-th iteration.\\
	\indent Problem \eqref{P1} is eventually transformed into a convex problem at the $(n+1)$-th iteration as follows 
	\begin{subequations}\label{P5}
		\begin{align}
			\mathop {\min }\limits_{ {\bm{a}}, {\bm{b}}, {\bm{\chi}}, t }
			& \quad -t+\lambda\hat{\varphi}({\bm{a}}, {\bm{b}}, {\bm{\chi}})
			\\
			\ \textrm{s.t.} \quad 
			&\eqref{C2},\eqref{C3},\eqref{C4},\eqref{P3b},\eqref{P3d},\eqref{P3e}, \eqref{P3_1_2}. \label{P5a}
		\end{align}
	\end{subequations}
	\subsection{Feasibility Check and Initialization}
	\indent We need to justify the feasibility of Problem \eqref{P1} first and find a feasible initial point for the proposed optimization algorithm. The feasibility check problem is formulated as follows
	\begin{subequations}\label{P6}
		\begin{align}
			\mathop {\text{find} }\quad
			& {\bm{a}}, {\bm{b}}
			\\
			\ \textrm{s.t.} \quad
			&R_{\mathrm{dl},k}\geq \kappa_{\mathrm{dl}},\forall k\in \mathcal{K}_{\mathrm{dl}},\label{P6a}\\
			&R_{\mathrm{ul},i}\geq \kappa_{\mathrm{ul}},\forall i\in \mathcal{K}_{\mathrm{ul}}, \label{P6b}\\
			& \eqref{P1c},\eqref{P1d}. \label{P6c}
		\end{align}
	\end{subequations}
	\indent By introducing the slack variable $\epsilon$, Problem \eqref{P6} can be relaxed as follows
	\begin{subequations}\label{P7}
		\begin{align}
			\mathop {\min }\limits_{ {\bm{a}}, {\bm{b}}, {\bm{\chi}}, \epsilon}
			&\epsilon+\lambda\hat{\varphi}({\bm{a}}, {\bm{b}}, {\bm{\chi}})
			\\
			\ \textrm{s.t.} \quad
			&R_{\mathrm{dl},k}+\epsilon\geq \kappa_{\mathrm{dl}},\forall k\in \mathcal{K}_{\mathrm{dl}},\label{P7a}\\
			& R_{\mathrm{ul},i}+\epsilon\geq \kappa_{\mathrm{ul}},\forall i\in \mathcal{K}_{\mathrm{ul}}, \label{P7b}\\
			& \eqref{P1c},\eqref{P3_1_2},\eqref{P3d},\eqref{P3e}.\label{P7f}
		\end{align}
	\end{subequations}
	\indent By using the following inequality \cite{vu2020cell}
	\begin{equation}
		\begin{aligned}
			\log(1+\frac{x^2}{y})&\geq \log(1+\frac{(x^{(n)})^2}{y^{(n)}})-\frac{(x^{(n)})^2}{y^{(n)}}\\&+2\frac{x^{(n)}x}{y^{(n)}}-\frac{(x^{(n)})^2(x^2+y)}{y^{(n)}((x^{(n)})^2+y^{(n)})},
		\end{aligned}
	\end{equation}
	we can obtain the lower bound of the DL rate and the UL rate as follows
	\begin{equation}
		\begin{aligned}
			R_{\mathrm{dl},k}\geq\widetilde{R}_{\mathrm{dl},k}&=\frac{\overline{\tau}}{\log2}\Big( \log(1+\frac{(\Psi_k^{(n)})^2}{\Omega_k^{(n)}})-\frac{(\Psi_k^{(n)})^2}{\Omega_k^{(n)}}\\&+2\frac{\Psi_k^{(n)}\Psi_k}{\Omega_k^{(n)}}-\frac{(\Psi_k^{(n)})^2(\Psi_k^2+\Omega_k)}{\Omega_k^{(n)}((\Psi_k^{(n)})^2+\Omega_k^{(n)})}\Big),
		\end{aligned}
	\end{equation}
	\begin{equation}
		\begin{aligned}
			R_{\mathrm{ul},i}\geq\widetilde{R}_{\mathrm{ul},i}&=\frac{\overline{\tau}}{\log2}\Big(\log(1+\frac{(\Xi_i^{(n)})^2}{\Pi_i^{(n)}})-\frac{(\Xi_i^{(n)})^2}{\Pi_i^{(n)}}\\&+2\frac{\Xi_i^{(n)}\Xi_i}{\Pi_i^{(n)}}-\frac{(\Xi_i^{(n)})^2(\Xi_i^2+\Pi_i)}{\Pi_i^{(n)}((\Xi_i^{(n)})^2+\Pi_i^{(n)})} \Big),
		\end{aligned}
	\end{equation}
	where $\Psi_k=\displaystyle\sum_{m=1}^{M}c_{\mathrm{dl},m,k}^{(1)}a_m$, $\Omega_k=\displaystyle\sum_{m=1}^{M}c_{\mathrm{dl},m,k}^{(2)}a_m+c_{\mathrm{dl},k}^{(3)}$, $\Xi_i=\displaystyle\sum_{m=1}^{M}c_{\mathrm{ul},m,i}^{(1)}b_m$, $\Pi_i=\displaystyle\sum_{m=1}^{M}c_{\mathrm{ul},m,i}^{(2)}b_m+\displaystyle\sum_{m=1}^{M}\displaystyle\sum_{l=1}^{M}c_{\mathrm{ul},m,l,i}^{(3)}\chi_{m,l}$.\\
	\indent Then, Problem \eqref{P7} can be transformed into a convex one as follows
	\begin{subequations}\label{P8}
		\begin{align}
			\mathop {\min }\limits_{ {\bm{a}}, {\bm{b}}, {\bm{\chi}}, \epsilon}
			&\epsilon+\lambda\varphi({\bm{a}}, {\bm{b}}, {\bm{\chi}})
			\\
			\ \textrm{s.t.} \quad
			&\widetilde{R}_{\mathrm{dl},k}+\epsilon\geq \kappa_{\mathrm{dl}},\forall k\in \mathcal{K}_{\mathrm{dl}},\label{P8a}\\
			& \widetilde{R}_{\mathrm{ul},i}+\epsilon\geq \kappa_{\mathrm{ul}},\forall i\in \mathcal{K}_{\mathrm{ul}}, \label{P8b}\\
			&\eqref{P3b},\eqref{P3_1_2},\eqref{P3d},\eqref{P3e}. \label{P8c}
		\end{align}
	\end{subequations}
	\indent We start with a random point and solve Problem \eqref{P8}. If the optimal $\epsilon$ is smaller than zero, Problem \eqref{P1} is feasible with constraints \eqref{P6a} and \eqref{P6b} satisfied.
	\subsection{Overall Algorithm and Complexity Analysis}
	\indent The detailed algorithm is presented in Algorithm 1. For Problem \eqref{P5}, the number of variables is $n_1=M^2+2M+1$. Constraint \eqref{C2} can be transformed into an equivalent quadratic constraint, and the other constraints are both linear constraints. Thus, Problem \eqref{P5} contains $K_{\mathrm{t}}$ quadratic constraints  and $n_2=K_{\mathrm{dl}}+K_{\mathrm{ul}}+5M^2+6M$ linear constraints. Consequently, the overall complexity of Problem \eqref{P5} is given by $\mathcal{O}\left(n_1^2\left(n_2+K_{\mathrm{t}}\right)^{\frac{1}{2}}\left(n_1+n_2+K_{\mathrm{t}}\right)\right)$ \cite{mohammadi2023network}.
%	The computational complexity of solving Problem \eqref{P5} mainly lies in the interior point method and is given by \cite{zhou2020framework}\\
%	\begin{equation}
%		{\footnotesize 
%			\mathcal{O}\left( \underbrace{\left(\sum_{j=1}^{J} k_{j}+2 m\right)^{\frac{1}{2}}}_{\text {Iteration complexity }}n(n^{2}+\underbrace{n \sum_{j=1}^{J} k_{j}^{2}+\sum_{j=1}^{J} k_{j}^{3}}_{\text {due to LMI }}+\underbrace{n\sum_{i=1}^{m} e_{i}^{2}}_{\text {due to SOC }})\right), }
%	\end{equation}
%	where $n$ denotes the number of variables, $J$ denotes the number of the linear matrix inequality (LMI) constraints, $k_{j}$ is the dimension of the $j$-th LMI constraint, $m$ denotes the number of the second-order cone (SOC) constraints, and $e_i$ is the dimension of the $i$-th SOC constraint.\\
	\begin{algorithm}
		\caption{Proposed Optimization Algorithm of AP Mode Selection}\label{A1}
		\begin{algorithmic}[1]
			\STATE Initialize iterative number $n=1$, maximum number of iterations $n_{\rm{max}}$,  feasible ${\bm{a}}^{(1)}$,
			${\bm{b}}^{(1)}$, ${\bm{\chi}}^{(1)}$ by solving Problem \eqref{P8}, error tolerance $\varepsilon$, calculate the OF value of Problem \eqref{P5}, denoted as ${\hat{f}}(t^{(1)},{{\bm{a}}^{(1)}},{\bm{b}}^{(1)},{\bm{\chi}}^{(1)}) \triangleq {\hat{f}}^{(1)}$;
			\STATE Given ${\bm{a}}^{(n)}$ and ${\bm{b}}^{(n)}$, ${\bm{\chi}}^{(n)}$ update the auxiliary variable ${\bm{\mu}}^{(n)}$ according to equation \eqref{80};
			\STATE Given ${\bm{\mu}}^{(n)}$, update $t^{(n+1)}$ ${\bm{a}}^{(n+1)}$,
			${\bm{b}}^{(n+1)}$, ${\bm{\chi}}^{(n+1)}$ by solving Problem \eqref{P8};
			\STATE If $n\geq n_{\rm{max}}$ or ${\left| {\hat{f}}^{(n+1)}- {\hat{f}}^{(n)}\right|} < \varepsilon$, terminate.  Otherwise, set $n \leftarrow n + 1$  and go to step 2.
		\end{algorithmic}
	\end{algorithm}

	\section{Simulation Results}\label{simulation}
	\subsection{Simulation Setup}
	\indent We consider a cooperative bi-static ISAC network where all the APs, UEs and targets are randomly distributed in a square of 0.3 $\times$ 0.3 $\mathrm{km}^2$, whose edges are wrapped around to avoid the boundary effects. The distances between the adjacent APs are at least 50 m \cite{bjornson2019making}. Some key parameters of the network are: $K_{\mathrm{dl}}=K_{\mathrm{ul}}=K_{\mathrm{t}}=4$, $\mathrm{RCS}=1$ $\mathrm{m}^2$, $\delta_{m,k}=2$. The number of symbols in each channel coherence time interval is $\tau_{c}=196$ \cite{zhang2014power}, $\tau_{\mathrm{ul}}=K_{\mathrm{dl}}+K_{\mathrm{ul}}$ symbols are utilized for UL channel estimation, and $\tau_{\mathrm{dl}}=M$ symbols are utilized for DL channel estimation. We further set the system frequency $f=4.8$ \text{GHz}, channel bandwidth $B=50$ \text{MHz} and the noise power density of -174 dBm/Hz. The transmit power of each UL UE is $p_{\mathrm{ul}}=0.1$ W and the total transmit power of each AP is 10 W, which is equally distributed to each DL UE and target.\\
	\indent The large-scale passloss of communication and radar in dB is respectively given by \eqref{PLc} and \eqref{PLs} as follows
	\begin{equation}\label{PLc}
		\begin{aligned}
			\mathrm{PL}_c=32.4+20\log_{10}(f)+20\log_{10}(d),
		\end{aligned}
	\end{equation}
	\begin{equation}\label{PLs}
		\begin{aligned}
			\mathrm{PL}_r&=103.4+20\log_{10}(f)+20\log_{10}(d_1d_2)-10\log_{10}(\mathrm{RCS}),
		\end{aligned}
	\end{equation}
	where $d$ denotes the communication link distance, $d_1$ and $d_2$ denote the distance from the target to the DL AP and the UL AP, respectively.
	
	\subsection{Benchmark}
	\indent In the simulation, we mainly consider four AP mode selection strategies as follows
	\begin{itemize}
		\item[$\bullet$] \textbf{Optimized mode}: The AP mode selection is determined by solving Problem \eqref{P5} according to Algorithm 1.
		\item[$\bullet$] \textbf{Greedy mode}: We design a sensing-centric greedy mode selection strategy based on the distance between the APs and the targets. Firstly, all the APs are set to DL mode. Then the distances between each AP and all the targets are calculated and the maximum distance is selected as the reference distance. The mode of the AP with the minimum reference distance is changed to UL mode. This process is repeated until the communication QoS constraints are satisfied and the minimum sensing SINR starts to decrease. The detailed procedure is presented in Algorithm 2.
		\item[$\bullet$] \textbf{The upper bound}: The AP mode selection is determined by solving Problem \eqref{P5} without the communication QoS constraints. Since the feasible domain of Problem \eqref{P5} is enlarged, we obtain the upper bound of the minimum sensing SINR.
		\item[$\bullet$] \textbf{Random mode}: The mode of APs is selected randomly.
	\end{itemize}
		\begin{algorithm}
		\caption{Greedy Algorithm of AP Mode Selection}\label{A2}
		\begin{algorithmic}[1]
			\STATE Initialize $\mathcal{M}_{\mathrm{c}}=\{1,\dots,M\}$, $\mathcal{M}_{\mathrm{s}}=\varnothing$, e.g. $\bm{a}=[1,1,\dots,1]^{\mathrm{T}}$, $\bm{b}=[0,0,\dots,0]^{\mathrm{T}}$;
			\STATE Calculate the distance $d_{m,p}$ between the $m$-th AP and the $p$-th target, and denote the maximum distance between the $m$-th AP and all the targets as $\mathop {\max }\limits_{p\in \mathcal{K}_{\mathrm{t}}}\{d_{m,p}\}$;
			\STATE \textbf{repeat}:
			\STATE \quad Choose index $m^{*}=\mathop{\arg\min}\limits_{m\in\mathcal{M}_{\mathrm{c}}}\mathop {\max }\limits_{p\in \mathcal{K}_{\mathrm{t}}}\{d_{m,p}\}$ as the UL AP;
			\STATE \quad $\mathcal{M}_{\mathrm{c}}\leftarrow\mathcal{M}_{\mathrm{c}}\backslash\{m^{*}\}$, $\mathcal{M}_{\mathrm{s}}\leftarrow\mathcal{M}_{\mathrm{s}}\cup\{m^{*}\}$;
			\STATE \quad Calculate the DL rate $R_{\mathrm{dl},k}$, UL rate $R_{\mathrm{ul},i}$, and the minimum sensing SINR;
			\STATE \textbf{Until} $\mathop{\min}\limits_{i\in K_{\mathrm{ul}}}R_{\mathrm{ul},i}\geq \kappa_{\mathrm{ul}}$, $\mathop{\min}\limits_{k\in K_{\mathrm{dl}}}R_{\mathrm{dl},k}\geq \kappa_{\mathrm{dl}}$ and $\mathop {\min}\limits_{p\in\mathcal{K}_{\mathrm{t}}}\mathrm{SINR}_{\mathrm{s},p}$ starts to decrease.
		\end{algorithmic}
	\end{algorithm}
	\subsection{Verification of the Closed-form Expressions}
\begin{figure}[H]
		\centering
		\includegraphics[width=3.2in]{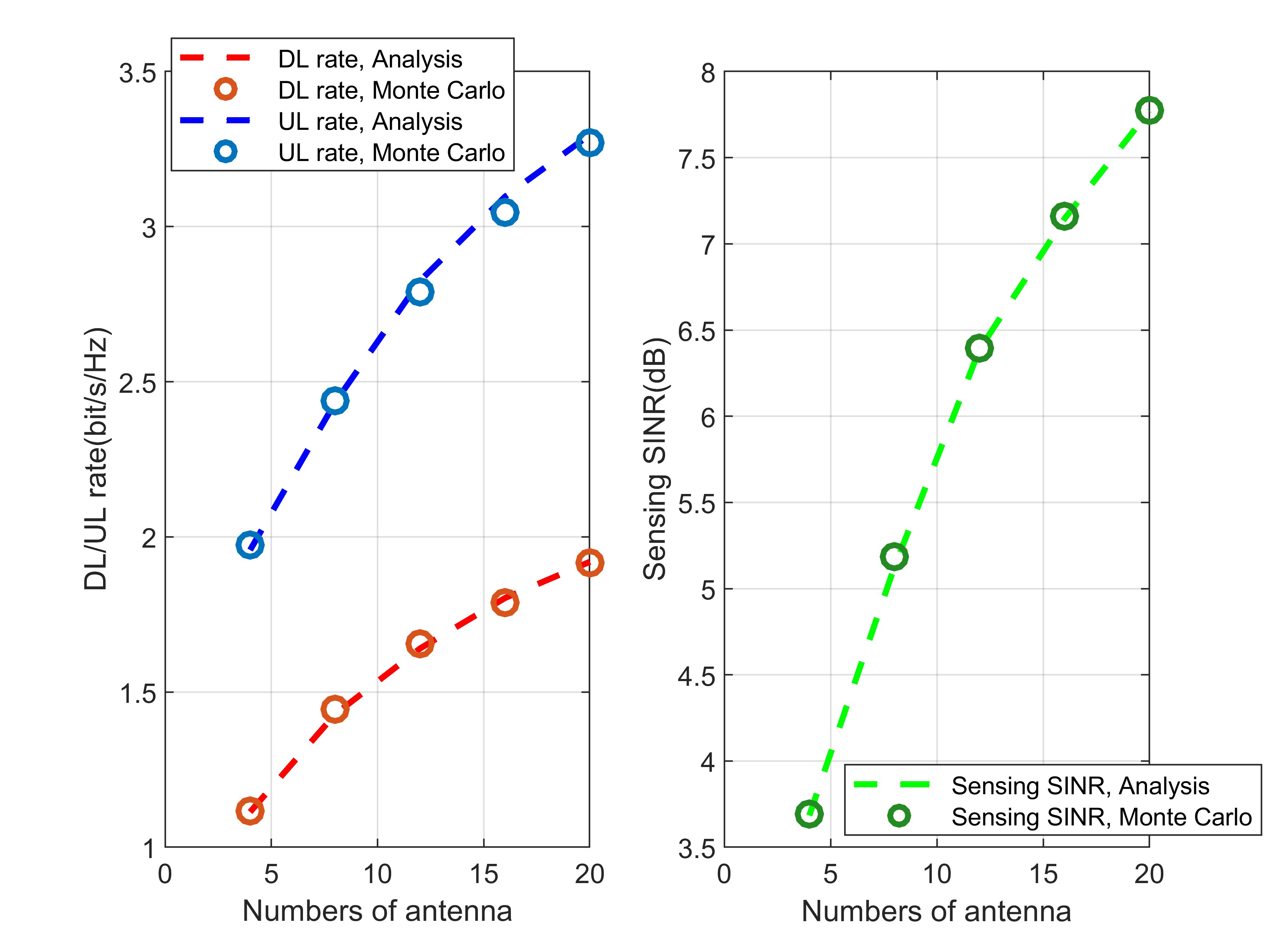}
		\caption{Verification of the closed-form expressions. (a) The closed-form expressions and Monte Carlo simulated values of the DL and UL rate versus the number of antennas. (b) The closed-form expressions and Monte Carlo simulated values of the sensing SINR versus the number of antennas.}
		\label{fig3}
	\end{figure}

	\indent As shown in Fig. \ref{fig3}, the results of the closed-form expressions are consistent with the Monte Carlo simulation results, and the consistency will not be influenced by changes in the number of antennas. This outcome confirms the accuracy of the derived closed-form expressions based on the long-term statistical CSI, which lays solid foundation for the implementation of the AP mode selection algorithm.
	
	\subsection{Convergence of the Proposed Optimization Algorithm}
\begin{figure}[H]
		\centering
		\includegraphics[width=3.2in]{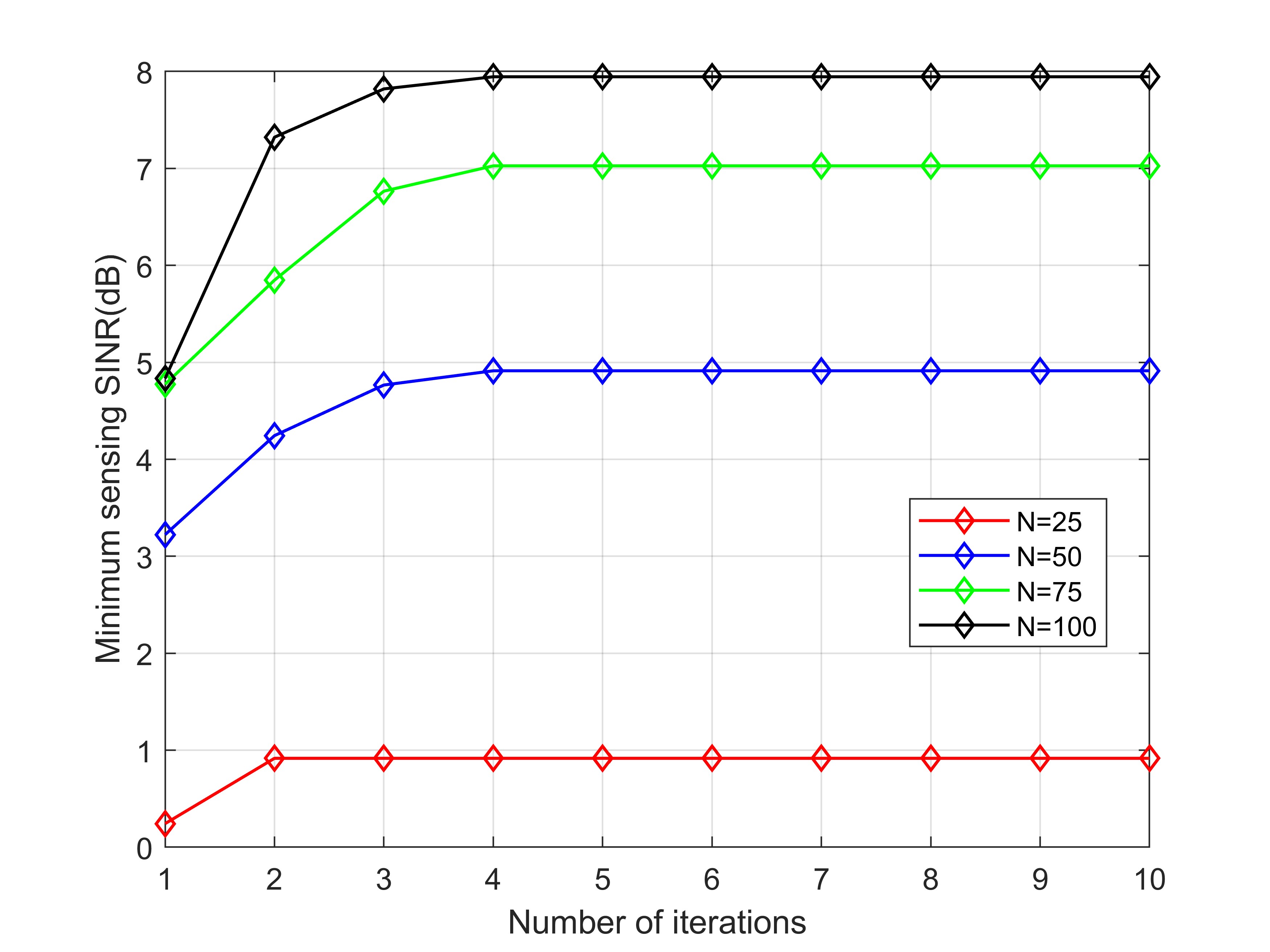}
		\caption{Convergence behavior of the proposed algorithm}
		\label{fig4}
	\end{figure}

	\indent We depict the trends of the minimum sensing SINR at each iteration in Fig. \ref{fig4} to validate the effectiveness and convergence of Algorithm 1 with $M=8$, $\kappa_{\mathrm{dl}}=\kappa_{\mathrm{ul}}=2$ bit/s/Hz. It is observed from Fig. \ref{fig4} that our proposed AP mode selection algorithm can achieve fast convergence within 5 iterations. Moreover, the convergence of Algorithm 1 is not affected by changes in the number of antennas.
	
	\subsection{Impact of Number of Antennas}
\begin{figure}[H]
		\centering
		\includegraphics[width=3.2in]{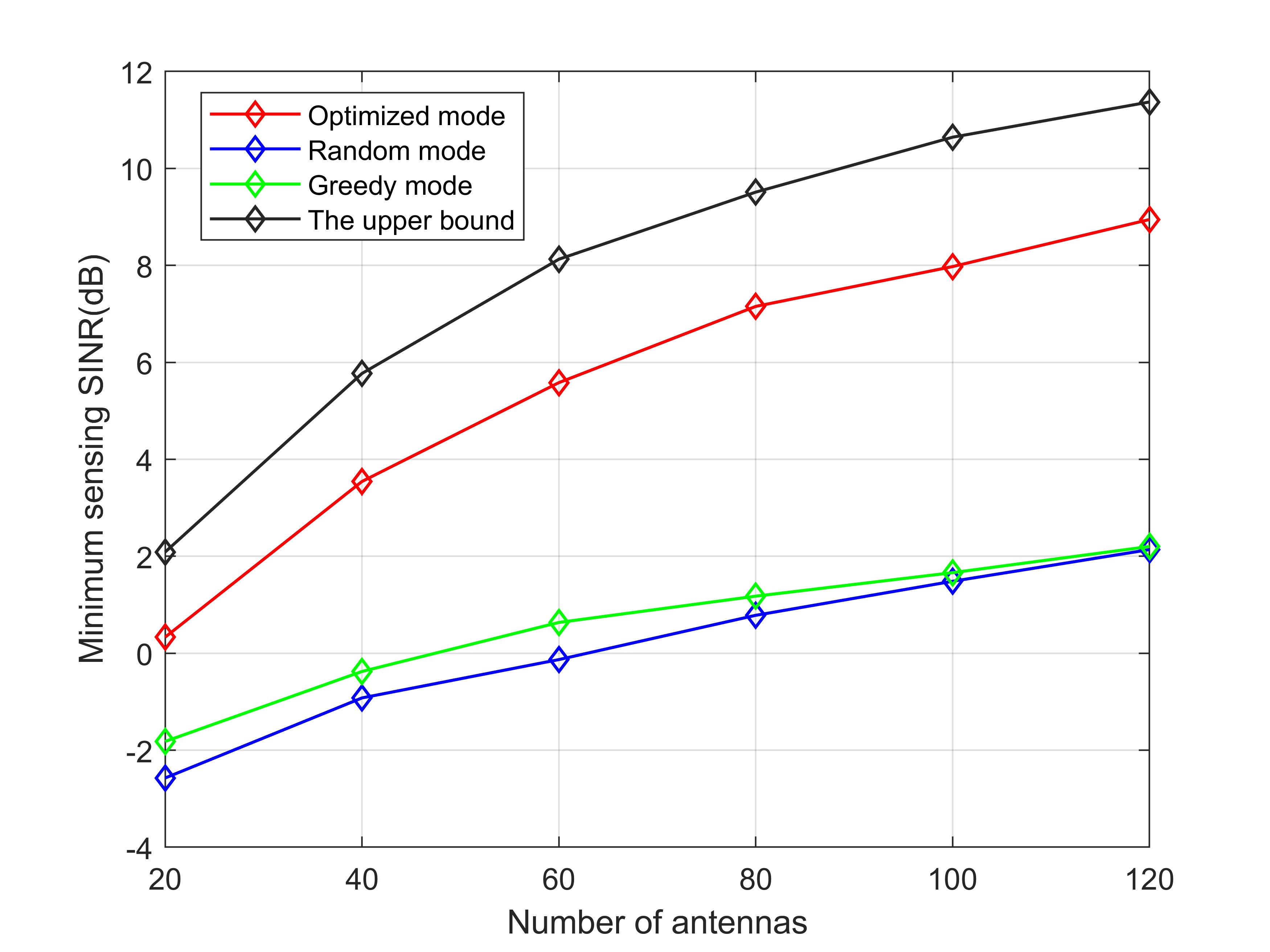}
		\caption{The minimum sensing SINR versus various numbers of antennas with $M=8$, $\kappa_{\mathrm{dl}}=\kappa_{\mathrm{ul}}=2$ bit/s/Hz.}
		\label{fig5}
	\end{figure}

	\indent Fig. \ref{fig5} illustrates the impact of number of antennas on the minimum sensing SINR under different AP mode selection strategies. It is explicitly shown that the minimum sensing SINR is monotonically enhanced when the number of antennas increases. Moreover, our proposed optimized mode can achieve better performance than the greedy mode and the random mode. Specifically, our proposed optimized mode can obtain 7 dB performance gain when $N=120$. However, our proposed optimized mode cannot achieve the upper bound because of the communication QoS constraints.
	
	\subsection{Impact of Number of APs}
\begin{figure}[H]
		\centering
		\includegraphics[width=3.2in]{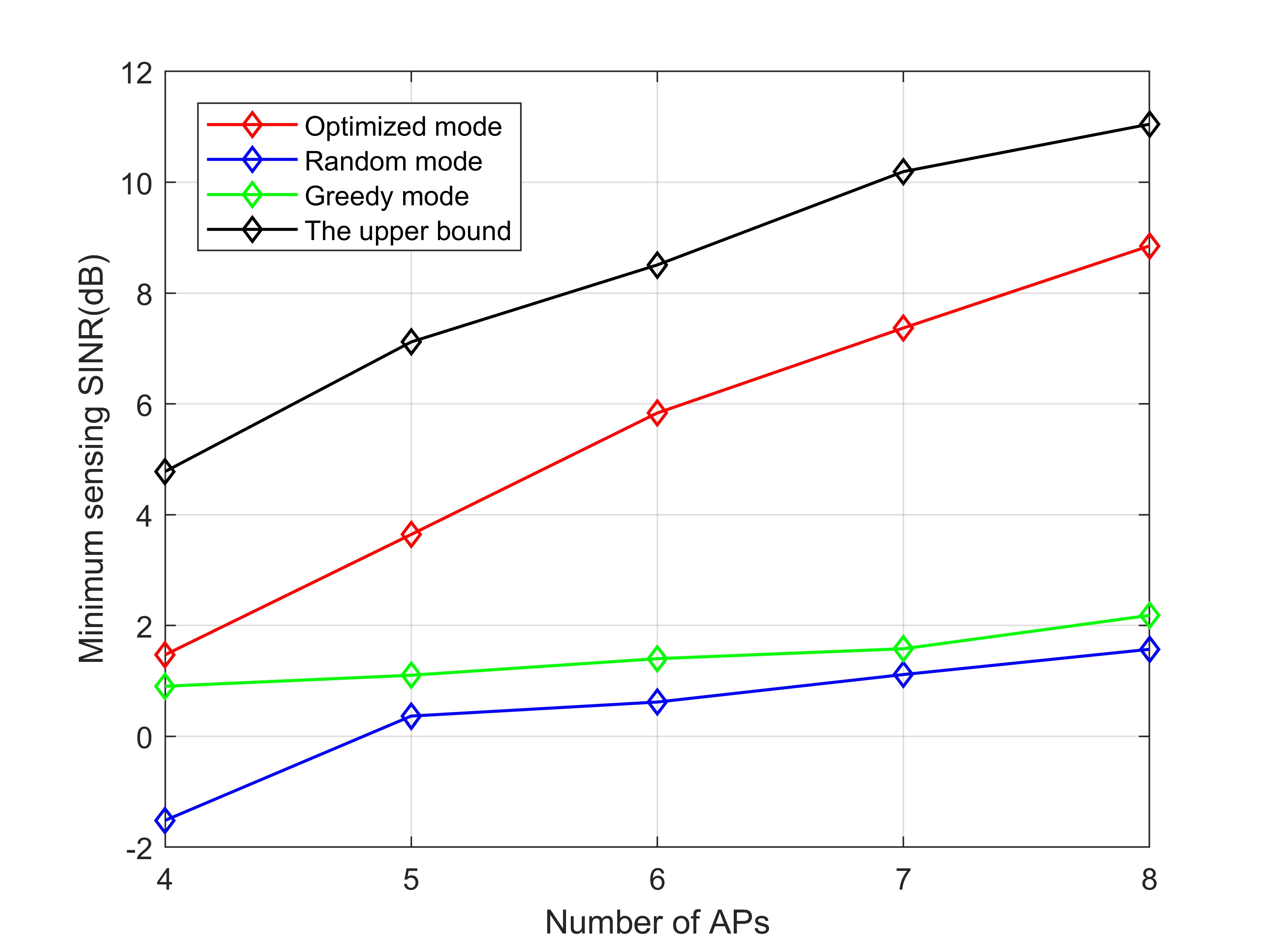}
		\caption{The minimum sensing SINR versus various numbers of APs with $N=100$, $\kappa_{\mathrm{dl}}=\kappa_{\mathrm{ul}}=2$ bit/s/Hz.}
		\label{fig6}
	\end{figure}

	\indent Fig. \ref{fig6} demonstrates the impact of number of APs on the minimum sensing SINR under different AP mode selection strategies. When the number of APs increase, despite the increase of the cross-link interference, the total power of the ISAC network increases, leading to the enhanced sensing performance. Furthermore, when the number of APs is small, the impact of the greedy mode can approach the optimized mode. However, when the number of APs continue to increase, our proposed optimized mode shows its unrivalled advantages.
	
	\subsection{Cumulative Distribution Function (CDF) of the Minimum Sensing SINR}
\begin{figure}[H]
		\centering
		\includegraphics[width=3.2in]{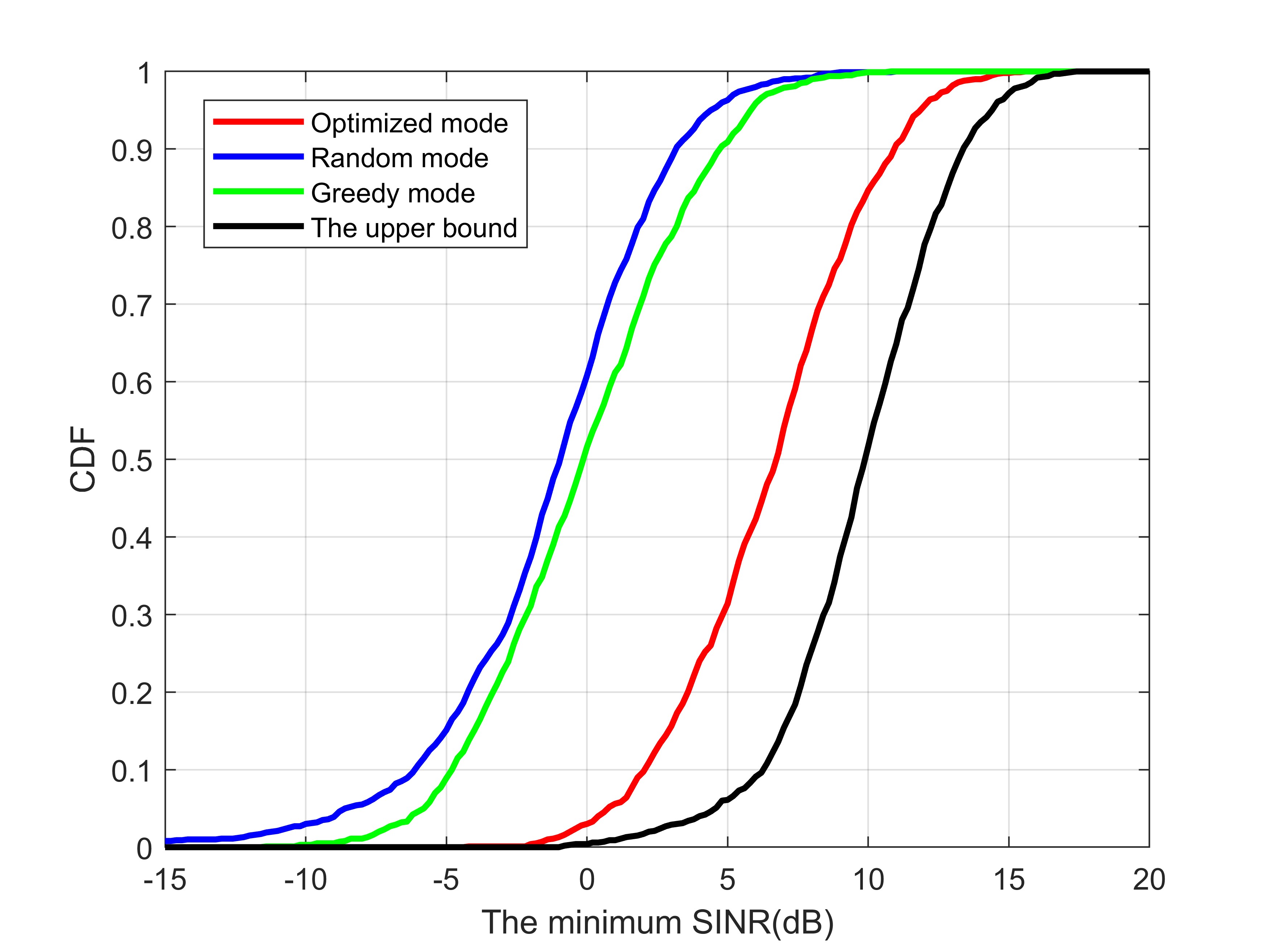}
		\caption{CDF of the minimum sensing SINR with $M=8$, $N=100$, $\kappa_{\mathrm{dl}}=\kappa_{\mathrm{ul}}=2$ bit/s/Hz.}
		\label{fig7}
	\end{figure}

	\indent Fig. \ref{fig7} presents the simulated CDF of the minimum sensing SINR under different mode selection strategies. It is seen that the performance gap between the optimized mode and the random mode is significant. When the value of CDF is 0.5, the minimum sensing SINR of the random mode, greedy mode, optimized mode and the upper bound is -0.98 dB, -0.16 dB, 6.72 dB and 9.94 dB, unveiling more than 6.5 dB performance gain of the optimized mode compared with the random mode and the greedy mode.
	
	\subsection{Impact of Communication QoS Threshold}
\begin{figure}[H]
		\centering
		\includegraphics[width=3.2in]{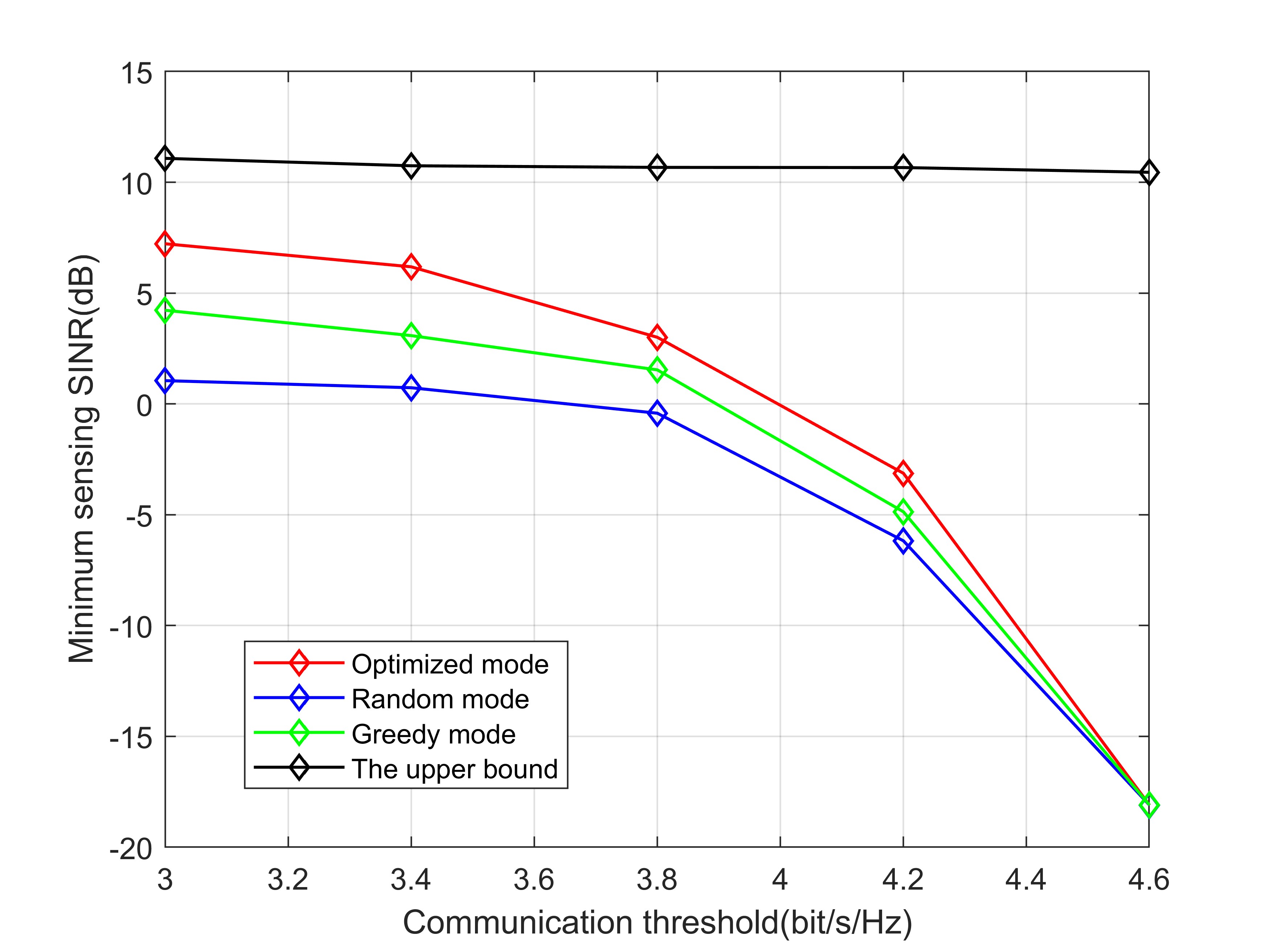}
		\caption{The minimum sensing SINR versus various communication QoS threshold with $M=8$, $N=100$.}
		\label{fig8}
	\end{figure}

	\indent We investigate the impact of communication QoS threshold on the minimum sensing SINR under different mode selection strategies. It is depicted in Fig. \ref{fig8} that the minimum sensing SINR decreases with the increase of the communication threshold, which reveals the trade-off between sensing and communication performance. It is worth noting that the minimum sensing SINR decreases to -17 dB when $\kappa_{\mathrm{dl}}=\kappa_{\mathrm{ul}}=4.6$ bit/s/Hz, because Problem \eqref{P5} becomes nearly infeasible under this circumstance.
	\section{Conclusion}\label{conclusion}
	\indent In this paper, we investigated the two-timescale AP mode selection design for the cooperative bi-static ISAC network. Firstly, the MMSE estimator was applied to estimate the channel between the APs and the channel from the APs to the UEs. Then, we adopted the low-complexity MRT beamforming and the MRC detector, and derived the closed-form expressions of the DL communication rate, the UL communication rate, and the sensing SINR. To obtain the trade-off between communication and sensing performance, we formulated a non-convex integer optimization problem to maximize the minimum sensing SINR under the communication QoS constraints. McCormick envelope relaxation and SCA techniques were applied to transform the NP-hard problem into more tractable forms and tackle the non-convexity. Tightness between the closed-form expressions and Monte Carlo simulations was justified. Simulation results proved the convergence of the proposed mode selection algorithm and its superior advantages over other benchmark schemes. Furthermore, the two-timescale design provided a practical AP mode selection strategy for the cooperative bi-static ISAC network with extremely low system overhead.

	\appendices
	\section{Some Useful Results}\label{A}
	\indent \emph{Lemma 1}: Consider a deterministic matrix $\mathbf{A} \in \mathbb{C}^{n\times n}$ and a matrix $\mathbf{X} \in \mathbb{C}^{m\times n}$ whose entries are i.i.d. random variables with zero mean and variance of $\sigma_x$. Then we have
	\begin{equation}
		\mathbb{E}\{\mathbf{X}\mathbf{A}\mathbf{X}^{\mathrm{H}}\}=\sigma_x\mathrm{Tr}\{\mathbf{A}\}\mathbf{I}_{m}.
	\end{equation}
	\indent \emph{Proof}: See \cite{zhi2022two}.\\
	\indent \emph{Lemma 2}: Consider a deterministic matrix $\mathbf{A} \in \mathbb{C}^{n\times n}$, deterministic vectors $\mathbf{w}_1, \mathbf{w}_2 \in \mathbb{C}^{n\times 1}$, and a vector $\mathbf{g} \in \mathbb{C}^{n\times 1}$ whose entries are i.i.d. random variables with zero mean and unit variance. Then we have
	\begin{equation}
		\mathbb{E}\{\mathbf{g}^{\mathrm{H}}\mathbf{w}_1\mathbf{g}^{\mathrm{H}}\mathbf{w}_2\}=\mathbb{E}\{\mathrm{Re}\{\mathbf{g}^{\mathrm{H}}\mathbf{w}_1\mathbf{g}^{\mathrm{H}}\mathbf{w}_2\}\}=0,
	\end{equation}
	\begin{equation}
		\mathbb{E}\{\mathbf{g}^{\mathrm{H}}\mathbf{A}\mathbf{g}\mathbf{g}^{\mathrm{H}}\mathbf{A}\mathbf{g}\}=\mathrm{Tr}\{\mathbf{A}^2\}+|\mathrm{Tr}\{\mathbf{A}\}|^2.
	\end{equation}
	\indent \emph{Proof}: See \cite{zhi2022two}.
	\section{Downlink Rate Derivation}\label{B}
	\indent Firstly, several closed-form expressions of expectations are calculated as follows
	\begin{equation}\label{B1}
		\begin{aligned}
			&1/\zeta_{m,k}^2=\mathbb{E}\{\left\| \hat{\mathbf{g}}_{m,k} \right\|_2^2\}=\tau_{\mathrm{ul}} p_{\mathrm{ul}}\Big(v_{m,k}\overline{\mathbf{g}}_{m,k}^{\mathrm{H}}\mathbf{A}_{m,k}^{\mathrm{H}}\mathbf{A}_{m,k}\overline{\mathbf{g}}_{m,k}\\&+u_{m,k}\mathrm{Tr}(\mathbf{A}_{m,i}^{\mathrm{H}}\mathbf{A}_{m,k})\Big)+\sigma^2\mathrm{Tr}(\mathbf{A}_{m,k}^{\mathrm{H}}\mathbf{A}_{m,k}),
		\end{aligned}
	\end{equation}
	\begin{equation}\label{B2}
		\begin{aligned}
			&\mathbb{E}\{\mathbf{g}_{m,k}^{\mathrm{H}}\hat{\mathbf{g}}_{m,k}\}=\mathbb{E}\{\sqrt{\tau_{\mathrm{ul}}p_{\mathrm{ul}}}\mathbf{g}_{m,k}^{\mathrm{H}}\mathbf{A}_{m,k}\mathbf{g}_{m,k}+\mathbf{g}_{m,k}^{\mathrm{H}}\mathbf{N}_m\boldsymbol{\varphi}_k\}\\&=\mathbb{E}\{\sqrt{\tau_{\mathrm{ul}}p_{\mathrm{ul}}}\mathbf{g}_{m,k}^{\mathrm{H}}\mathbf{A}_{m,k}\mathbf{g}_{m,k}\}=\sqrt{\tau_{\mathrm{ul}}p_{\mathrm{ul}}}(v_{m,k}\overline{\mathbf{g}}_{m,k}^{\mathrm{H}}\mathbf{A}_{m,k}\overline{\mathbf{g}}_{m,k}\\&+u_{m,k}\mathbb{E}\{\widetilde{\mathbf{g}}_{m,k}^{\mathrm{H}}\mathbf{A}_{m,k}\widetilde{\mathbf{g}}_{m,k}\})=\sqrt{\tau_{\mathrm{ul}}p_{\mathrm{ul}}}v_{m,k}\overline{\mathbf{g}}_{m,k}^{\mathrm{H}}\mathbf{A}_{m,k}\overline{\mathbf{g}}_{m,k}\\&+\sqrt{\tau_{\mathrm{ul}}p_{\mathrm{ul}}}u_{m,k}\mathrm{Tr}(\mathbf{A}_{m,k}),
		\end{aligned}
	\end{equation}
	\begin{figure*}[t]
		\begin{equation}\label{BU}
			\begin{aligned}
				&\mathbb{E}\{|\mathbf{g}_{m,k}^{\mathrm{H}}\hat{\mathbf{g}}_{m,k}|^2\}=\mathbb{E}\{|\sqrt{\tau_{\mathrm{ul}}p_{\mathrm{ul}}}\mathbf{g}_{m,k}^{\mathrm{H}}\mathbf{A}_{m,k}\mathbf{g}_{m,k}+\mathbf{g}_{m,k}^{\mathrm{H}}\mathbf{A}_{m,k}\mathbf{N}_m\boldsymbol{\varphi}_k|^2\}\\&=\tau_{\mathrm{ul}}p_{\mathrm{ul}}\mathbb{E}\{\mathbf{g}_{m,k}^{\mathrm{H}}\mathbf{A}_{m,k}\mathbf{g}_{m,k}\mathbf{g}_{m,k}^{\mathrm{H}}\mathbf{A}_{m,k}^{\mathrm{H}}\mathbf{g}_{m,k}\}+\mathbb{E}\{\mathbf{g}_{m,k}^{\mathrm{H}}\mathbf{A}_{m,k}\mathbf{N}_m\mathbf{N}_m^{\mathrm{H}}\mathbf{A}_{m,k}^{\mathrm{H}}\mathbf{g}_{m,k}\}\\&=\tau_{\mathrm{ul}}p_{\mathrm{ul}}\Big(v_{m,k}^2\overline{\mathbf{g}}_{m,k}^{\mathrm{H}}\mathbf{A}_{m,k}\overline{\mathbf{g}}_{m,k}\overline{\mathbf{g}}_{m,k}^{\mathrm{H}}\mathbf{A}_{m,k}^{\mathrm{H}}\overline{\mathbf{g}}_{m,k}+\mathbb{E}\{u_{m,k}v_{m,k}\widetilde{\mathbf{g}}_{m,k}^{\mathrm{H}}\mathbf{A}_{m,k}\overline{\mathbf{g}}_{m,k}\overline{\mathbf{g}}_{m,k}^{\mathrm{H}}\mathbf{A}_{m,k}^{\mathrm{H}}\widetilde{\mathbf{g}}_{m,k}\}\\&+\mathbb{E}\{u_{m,k}v_{m,k}\overline{\mathbf{g}}_{m,k}^{\mathrm{H}}\mathbf{A}_{m,k}\overline{\mathbf{g}}_{m,k}\widetilde{\mathbf{g}}_{m,k}^{\mathrm{H}}\mathbf{A}_{m,k}^{\mathrm{H}}\widetilde{\mathbf{g}}_{m,k}\}+\mathbb{E}\{u_{m,k}v_{m,k}\widetilde{\mathbf{g}}_{m,k}^{\mathrm{H}}\mathbf{A}_{m,k}\overline{\mathbf{g}}_{m,k}\widetilde{\mathbf{g}}_{m,k}^{\mathrm{H}}\mathbf{A}_{m,k}^{\mathrm{H}}\overline{\mathbf{g}}_{m,k}\}\\&+\mathbb{E}\{u_{m,k}v_{m,k}\overline{\mathbf{g}}_{m,k}^{\mathrm{H}}\mathbf{A}_{m,k}\widetilde{\mathbf{g}}_{m,k}\overline{\mathbf{g}}_{m,k}^{\mathrm{H}}\mathbf{A}_{m,k}^{\mathrm{H}}\widetilde{\mathbf{g}}_{m,k}\}+\mathbb{E}\{u_{m,k}v_{m,k}\widetilde{\mathbf{g}}_{m,k}^{\mathrm{H}}\mathbf{A}_{m,k}\widetilde{\mathbf{g}}_{m,k}\overline{\mathbf{g}}_{m,k}^{\mathrm{H}}\mathbf{A}_{m,k}^{\mathrm{H}}\overline{\mathbf{g}}_{m,k}\}\\&+\mathbb{E}\{u_{m,k}v_{m,k}\overline{\mathbf{g}}_{m,k}^{\mathrm{H}}\mathbf{A}_{m,k}\widetilde{\mathbf{g}}_{m,k}\widetilde{\mathbf{g}}_{m,k}^{\mathrm{H}}\mathbf{A}_{m,k}^{\mathrm{H}}\overline{\mathbf{g}}_{m,k}\}+\mathbb{E}\{u_{m,k}^2\widetilde{\mathbf{g}}_{m,k}^{\mathrm{H}}\mathbf{A}_{m,k}\widetilde{\mathbf{g}}_{m,k}\widetilde{\mathbf{g}}_{m,k}^{\mathrm{H}}\mathbf{A}_{m,k}^{\mathrm{H}}\widetilde{\mathbf{g}}_{m,k}\}\Big)+\sigma^2\mathbb{E}\{\mathbf{g}_{m,k}^{\mathrm{H}}\mathbf{A}_{m,k}\mathbf{A}_{m,k}^{\mathrm{H}}\mathbf{g}_{m,k}\}\\&=\tau_{\mathrm{ul}}p_{\mathrm{ul}}v_{m,k}^2\overline{\mathbf{g}}_{m,k}^{\mathrm{H}}\mathbf{A}_{m,k}\overline{\mathbf{g}}_{m,k}\overline{\mathbf{g}}_{m,k}^{\mathrm{H}}\mathbf{A}_{m,k}^{\mathrm{H}}\overline{\mathbf{g}}_{m,k}+2\tau_{\mathrm{ul}}p_{\mathrm{ul}}u_{m,k}v_{m,k}\mathrm{Tr}\{\mathbf{A}_{m,k}\overline{\mathbf{g}}_{m,k}\overline{\mathbf{g}}_{m,k}^{\mathrm{H}}\mathbf{A}_{m,k}^{\mathrm{H}}\}\\&+2\tau_{\mathrm{ul}}p_{\mathrm{ul}}u_{m,k}v_{m,k}\overline{\mathbf{g}}_{m,k}^{\mathrm{H}}\mathbf{A}_{m,k}\overline{\mathbf{g}}_{m,k}\mathrm{Tr}\{\mathbf{A}_{m,k}\}+\tau_{\mathrm{ul}}p_{\mathrm{ul}}u_{m,k}^2(\mathrm{Tr}\{\mathbf{A}_{m,k}^2\}+|\mathrm{Tr}\{\mathbf{A}_{m,k}\}|^2),
			\end{aligned}
		\end{equation}
		\centering
		\vspace*{0pt}
		\hrulefill
		\vspace*{0pt} 
	\end{figure*}
	\begin{equation}
		\mathbb{E}\{|\sqrt{p_{\mathrm{ul}}}h_{k,j}|^2\}=p_{\mathrm{ul}}\alpha_{k,j}.
	\end{equation}
	\indent for $i \neq k$, we have
	\begin{equation}\label{B3}
		\begin{aligned}
			&\mathbb{E}\{|\mathbf{g}_{m,k}^{\mathrm{H}}\hat{\mathbf{g}}_{m,i}|^2\}\\&=\mathbb{E}\{|\sqrt{\tau_{\mathrm{ul}}p_{\mathrm{ul}}}\mathbf{g}_{m,k}^{\mathrm{H}}\mathbf{A}_{m,i}\mathbf{g}_{m,i}+\mathbf{g}_{m,k}^{\mathrm{H}}\mathbf{A}_{m,i}\mathbf{N}_m\boldsymbol{\varphi}_i|^2\}\\&=\tau_{\mathrm{ul}}p_{\mathrm{ul}}\mathbb{E}\{\mathbf{g}_{m,k}^{\mathrm{H}}\mathbf{A}_{m,i}\mathbb{E}\{\mathbf{g}_{m,i}\mathbf{g}_{m,i}^{\mathrm{H}}\}\mathbf{A}_{m,i}^{\mathrm{H}}\mathbf{g}_{m,k}\}\\&+\mathbb{E}\{\mathbf{g}_{m,k}^{\mathrm{H}}\mathbf{A}_{m,i}\mathbf{N}_m\mathbf{N}_m^{\mathrm{H}}\mathbf{A}_{m,i}^{\mathrm{H}}\mathbf{g}_{m,k}\}=\mathbb{E}\{\mathbf{g}_{m,k}^{\mathrm{H}}\mathbf{B}_{m,i}\mathbf{g}_{m,k}\}\\&=v_{m,k}\overline{\mathbf{g}}_{m,k}^{\mathrm{H}}\mathbf{B}_{m,i}\overline{\mathbf{g}}_{m,k}+u_{m,k}\mathrm{Tr}(\mathbf{B}_{m,i}).
		\end{aligned}
	\end{equation}
	%where $\mathbf{B}_{m,i}=\mathbf{A}_{m,i}\left(\tau_{\mathrm{ul}}p_{\mathrm{ul}}v_{m,i}\overline{\mathbf{g}}_{m,i}\overline{\mathbf{g}}_{m,i}^{\mathrm{H}}+\left(\tau_{\mathrm{ul}}p_{\mathrm{ul}}u_{m,i}+\sigma^2\right)\mathbf{I}_N\right)\mathbf{A}_{m,i}^{\mathrm{H}}$.
	By letting $\mathbf{R}_{m}=\displaystyle\sum_{t=1}^{K_{\mathrm{t}}}p_{\mathrm{s},m,t}\mathbf{f}_{m,t}\mathbf{f}_{m,t}^{\mathrm{H}}$, we have
	\begin{equation}\label{B4}
		\mathbb{E}\{\mathbf{g}_{m,k}^{\mathrm{H}}\mathbf{R}_m\mathbf{g}_{m,k}\}=v_{m,k}\overline{\mathbf{g}}_{m,k}^{\mathrm{H}}\mathbf{R}_m\overline{\mathbf{g}}_{m,k}+u_{m,k}\mathrm{Tr}(\mathbf{R}_m).
	\end{equation}
	\indent The derivations of \eqref{B1}, \eqref{B2}, \eqref{BU}, \eqref{B3}, \eqref{B4} are completed by using Lemma 1, and the derivation of \eqref{BU} also uses Lemma 2 for $\widetilde{\mathbf{g}}_{m,k}\sim \mathcal{CN}(\mathbf{0},\mathbf{I}_N)$.\\
	\indent As shown in \eqref{dl}, the strength of the signal received by the $k$-th DL UE consists of the desired DL signal $E_{\mathrm{dl},k}$, the beamforming gain uncertainty $N_{\mathrm{dl},k}^{\mathrm{BU}}$, the cross-link interference $N_{\mathrm{dl},k}^{\mathrm{MUI}}$ caused by the other DL UEs, the cross-link interference $N_{\mathrm{dl},k}^{\mathrm{UL}}$ caused by the UL UEs and the sensing symbol interference $	N_{\mathrm{dl},k}^{\mathrm{S}}$, which is respectively calculated as follows
	\begin{equation}
		\begin{aligned}
			E_{\mathrm{dl},k}&=\left(\displaystyle\sum_{m=1}^{M}a_m\zeta_{m,k}\sqrt{p_{\mathrm{dl},m,k}}\mathbb{E}\{\mathbf{g}_{m,k}^{\mathrm{H}}\hat{\mathbf{g}}_{m,k}\}\right)^2\\&\triangleq \left(\displaystyle\sum_{m=1}^{M}c_{\mathrm{dl},m,k}^{(1)}a_m\right)^2,
		\end{aligned}
	\end{equation}
	\begin{equation}
		\begin{aligned}
			N_{\mathrm{dl},k}^{\mathrm{BU}}&=\displaystyle\sum_{m=1}^{M}a_m^2\zeta_{m,k}^2p_{\mathrm{dl},m,k}(\mathbb{E}\{|\mathbf{g}_{m,k}^{\mathrm{H}}\hat{\mathbf{g}}_{m,k}|^2\}\\&-|\mathbb{E}\{\mathbf{g}_{m,k}^{\mathrm{H}}\hat{\mathbf{g}}_{m,k}\}|^2) \triangleq \displaystyle\sum_{m=1}^{M}c_{\mathrm{dl},m,k}^{(\mathrm{I},1)}a_m^2,
		\end{aligned}
	\end{equation}
	\begin{equation}
		\begin{aligned}
			N_{\mathrm{dl},k}^{\mathrm{MUI}}&=\displaystyle\sum_{m=1}^{M}\displaystyle\sum_{i\neq k}a_m^2\zeta_{m,i}^2p_{\mathrm{dl},m,i}	\mathbb{E}\{|\mathbf{g}_{m,k}^{\mathrm{H}}\hat{\mathbf{g}}_{m,i}|^2\}\\&\triangleq \displaystyle\sum_{m=1}^{M}c_{\mathrm{dl},m,k}^{(\mathrm{I},2)}a_m^2,
		\end{aligned}
	\end{equation}
	\begin{equation}
		N_{\mathrm{dl},k}^{\mathrm{UL}}=\displaystyle\sum_{j\in\mathcal{K}_{\mathrm{ul}}}\mathbb{E}\{|\sqrt{p_{\mathrm{ul}}}h_{k,j}|^2\}=\displaystyle\sum_{j\in\mathcal{K}_{\mathrm{ul}}}p_{\mathrm{ul}}\alpha_{k,j},
	\end{equation}
	\begin{equation}
		N_{\mathrm{dl},k}^{\mathrm{S}}=\displaystyle\sum_{m=1}^{M}a_m^2\mathbb{E}\{\mathbf{g}_{m,k}^{\mathrm{H}}\mathbf{R}_m\mathbf{g}_{m,k}\}\triangleq \displaystyle\sum_{m=1}^{M}c_{\mathrm{dl},m,k}^{(\mathrm{I},3)}a_m^2.
	\end{equation}
	\indent By letting $c_{\mathrm{dl},m,k}^{(2)}=c_{\mathrm{dl},m,k}^{(\mathrm{I},1)}+c_{\mathrm{dl},m,k}^{(\mathrm{I},2)}+c_{\mathrm{dl},m,k}^{(\mathrm{I},3)}$ and $c_{\mathrm{dl},k}^{(3)}=N_{\mathrm{dl},k}^{\mathrm{UL}}+\sigma^2$, the desired result in \eqref{dl_rate} is obtained.
	\section{Uplink Rate Derivation}\label{C}
	\indent Firstly, several closed-form expressions of expectations are calculated as follows
	\begin{equation}\label{U1}
		\begin{aligned}
			\mathbb{E}\{\hat{\mathbf{g}}_{m,i}\hat{\mathbf{g}}_{m,i}^{\mathrm{H}}\}&=\mathbb{E}\{\tau_{\mathrm{ul}}p_{\mathrm{ul}}\mathbf{A}_{m,i}\mathbf{g}_{m,i}\mathbf{g}_{m,i}^{\mathrm{H}}\mathbf{A}_{m,i}^{\mathrm{H}}\}\\&+\mathbb{E}\{\mathbf{A}_{m,i}\mathbf{N}_{m}\mathbf{N}_{m}^{\mathrm{H}}\mathbf{A}_{m,i}^{\mathrm{H}}\}=\mathbf{B}_{m,i},
		\end{aligned}
	\end{equation}
	\begin{equation}\label{U2}
		\begin{aligned}
			&\mathbb{E}\{|\hat{\mathbf{g}}_{m,i}^{\mathrm{H}}\mathbf{e}_{m,l}^{\mathrm{AP}}\hat{\mathbf{g}}_{l,k}|^2\}=\mathbb{E}\{\hat{\mathbf{g}}_{m,i}^{\mathrm{H}}\mathbf{e}_{m,l}^{\mathrm{AP}}\hat{\mathbf{g}}_{l,k}\hat{\mathbf{g}}_{l,k}^{\mathrm{H}}(\mathbf{e}_{m,l}^{\mathrm{AP}})^{\mathrm{H}}\hat{\mathbf{g}}_{m,i}\}\\&=\mathbb{E}\{\hat{\mathbf{g}}_{m,i}^{\mathrm{H}}\mathbf{e}_{m,l}^{\mathrm{AP}}\mathbb{E}\{\hat{\mathbf{g}}_{l,k}\hat{\mathbf{g}}_{l,k}^{\mathrm{H}}\}(\mathbf{e}_{m,l}^{\mathrm{AP}})^{\mathrm{H}}\hat{\mathbf{g}}_{m,i}\}\\&=\mathbb{E}\{\hat{\mathbf{g}}_{m,i}^{\mathrm{H}}\mathbf{e}_{m,l}^{\mathrm{AP}}\mathbf{B}_{l,k}(\mathbf{e}_{m,l}^{\mathrm{AP}})^{\mathrm{H}}\hat{\mathbf{g}}_{m,i}\}\\&=\frac{\sigma^2\gamma_{m,l}\mathrm{Tr}(\mathbf{B}_{l,k})\mathbb{E}\{\hat{\mathbf{g}}_{m,i}^{\mathrm{H}}\hat{\mathbf{g}}_{m,i}\}}{\tau_{\mathrm{dl}} p_{\mathrm{dl}}\gamma_{m,l}+\sigma^2}=\frac{\sigma^2\gamma_{m,l}\zeta_{m,i}^2}{\tau_{\mathrm{dl}} p_{\mathrm{dl}}\gamma_{m,l}+\sigma^2}\mathrm{Tr}(\mathbf{B}_{l,k}),
		\end{aligned}
	\end{equation}
	\begin{equation}\label{U3}
		\begin{aligned}
			&\mathbb{E}\{\hat{\mathbf{g}}_{m,i}^{\mathrm{H}}\mathbf{e}_{m,l}^{\mathrm{AP}}\mathbf{R}_{l}(\mathbf{e}_{m,l}^{\mathrm{AP}})^{\mathrm{H}}\hat{\mathbf{g}}_{m,i}\}=\frac{\sigma^2\gamma_{m,l}\mathrm{Tr}(\mathbf{R}_{l})\mathbb{E}\{\hat{\mathbf{g}}_{m,i}^{\mathrm{H}}\hat{\mathbf{g}}_{m,i}\}}{\tau_{\mathrm{dl}} p_{\mathrm{dl}}\gamma_{m,l}+\sigma^2}\\&=\frac{\sigma^2\gamma_{m,l}\zeta_{m,i}^2}{\tau_{\mathrm{dl}} p_{\mathrm{dl}}\gamma_{m,l}+\sigma^2}\mathrm{Tr}(\mathbf{R}_{l}).
		\end{aligned}
	\end{equation}
	\indent The derivations of \eqref{U1}, \eqref{U2}, \eqref{U3} are completed by using Lemma 1 for $\widetilde{\mathbf{g}}_{m,i}\sim \mathcal{CN}(\mathbf{0},\mathbf{I}_N)$ and $\mathbf{e}_{m,l}^{\mathrm{AP}}\sim \mathcal{CN}(\mathbf{0},\frac{\sigma^2\gamma_{m,l}}{\tau_{\mathrm{dl}} p_{\mathrm{dl}}\gamma_{m,l}+\sigma^2}\mathbf{I}_N)$.\\
	\indent For the UL data transmission, as shown in \eqref{ul}, the received signal strength after the MRC detector consists of the desired UL signal $E_{\mathrm{ul},i}$, the beamforming gain uncertainty $N_{\mathrm{ul},i}^{\mathrm{BU}}$, the cross-link interference $N_{\mathrm{ul},i}^{\mathrm{MUI}}$ caused by the other UL UEs, the residual interference $N_{\mathrm{ul},i}^{\mathrm{D}}$ from the DL APs and the echo interference $N_{\mathrm{ul},i}^{\mathrm{ECHO}}$, which is respectively calculated as
	\begin{equation}
		\begin{aligned}
			E_{\mathrm{ul},i}=\left(\displaystyle\sum_{m=1}^{M}b_m\sqrt{p_{\mathrm{ul}}}\mathbb{E}\{\hat{\mathbf{g}}_{m,i}^{\mathrm{H}}\mathbf{g}_{m,i}\}\right)^2\triangleq \left(\displaystyle\sum_{m=1}^{M}c_{\mathrm{ul},m,i}^{(1)}b_m\right)^2,
		\end{aligned}
	\end{equation}
	\begin{equation}
		\begin{aligned}
			N_{\mathrm{ul},i}^{\mathrm{BU}}&=\displaystyle\sum_{m=1}^{M}b_m^2p_{\mathrm{ul}}(\mathbb{E}\{|\hat{\mathbf{g}}_{m,i}^{\mathrm{H}}\mathbf{g}_{m,i}|^2\}-|\mathbb{E}\{\hat{\mathbf{g}}_{m,i}^{\mathrm{H}}\mathbf{g}_{m,i}\}|^2)\\&\triangleq \displaystyle\sum_{m=1}^{M}c_{\mathrm{ul},m,i}^{(\mathrm{I},1)}b_m^2,
		\end{aligned}
	\end{equation}
	\begin{equation}
		\begin{aligned}
			N_{\mathrm{ul},i}^{\mathrm{MUI}}=\displaystyle\sum_{m=1}^{M}\displaystyle\sum_{j\in\mathcal{K}\backslash\{i\}}b_m^2p_{\mathrm{ul}}\mathbb{E}\{|\hat{\mathbf{g}}_{m,i}^{\mathrm{H}}\mathbf{g}_{m,j}|^2\}\triangleq \displaystyle\sum_{m=1}^{M}c_{\mathrm{ul},m,i}^{(\mathrm{I},2)}b_m^2,
		\end{aligned}
	\end{equation}
	\begin{equation}
		\begin{aligned}
			N_{\mathrm{ul},i}^{\mathrm{D}}&=\displaystyle\sum_{m=1}^{M}\displaystyle\sum_{l=1}^{M}b_m^2a_l^2\Big(\displaystyle\sum_{k=1}^{K_{\mathrm{dl}}}p_{\mathrm{dl},l,k}\mathbb{E}\{|\hat{\mathbf{g}}_{m,i}^{\mathrm{H}}\mathbf{e}_{m,l}^{\mathrm{AP}}\hat{\mathbf{g}}_{l,k}|^2\}\\&+\mathbb{E}\{\hat{\mathbf{g}}_{m,i}^{\mathrm{H}}\mathbf{e}_{m,l}^{\mathrm{AP}}\mathbf{R}_l(\mathbf{e}_{m,l}^{\mathrm{AP}})^{\mathrm{H}}\hat{\mathbf{g}}_{m,i}\}\Big)\triangleq \displaystyle\sum_{m=1}^{M}\displaystyle\sum_{l=1}^{M}c_{\mathrm{ul},m,l,i}^{(\mathrm{I},3)}a_l^2b_m^2,
		\end{aligned}
	\end{equation}
	\begin{equation}
		\begin{aligned}
			N_{\mathrm{ul},i}^{\mathrm{ECHO}}&=\displaystyle\sum_{m=1}^{M}\displaystyle\sum_{t=1}^{K_{\mathrm{t}}}\displaystyle\sum_{l=1}^{M}a_l^2b_m^2\mathbb{E}\{|\hat{\mathbf{g}}_{m,i}^{\mathrm{H}}\mathbf{y}_{\mathrm{echo},m}|^2\}\\&\triangleq \displaystyle\sum_{m=1}^{M}\displaystyle\sum_{l=1}^{M}c_{\mathrm{ul},m,l,i}^{(\mathrm{I},4)}a_l^2b_m^2.
		\end{aligned}
	\end{equation}
	\indent By letting $c_{\mathrm{ul},m,i}^{(2)}=c_{\mathrm{ul},m,i}^{(\mathrm{I},1)}+c_{\mathrm{ul},m,i}^{(\mathrm{I},2)}+\sigma^2\zeta_{m,i}^2$ and $c_{\mathrm{ul},m,l,i}^{(3)}=c_{\mathrm{ul},m,l,i}^{(\mathrm{I},3)}+c_{\mathrm{ul},m,l,i}^{(\mathrm{I},4)}$, the desired result in \eqref{ul_rate} is obtained.
	\section{Sensing SINR Derivation}\label{D}
	\indent Firstly, two closed-form expressions of expectations are calculated as follows
	\begin{equation}\label{D1}
		\begin{aligned}
			&\mathbb{E}\{|\mathbf{u}_{m,p}^{\mathrm{H}}\sqrt{p_{\mathrm{ul}}}(\mathbf{g}_{m,j}-\hat{\mathbf{g}}_{m,j})|^2\}\\&=p_{\mathrm{ul}}\mathbf{u}_{m,p}^{\mathrm{H}}\mathbb{E}\{(\mathbf{e}_{m,j}^{\mathrm{C}})^{\mathrm{H}}\mathbf{e}_{m,j}^{\mathrm{C}}\}\mathbf{u}_{m,p}=p_{\mathrm{ul}}N^2u_{m,j}\mathrm{NMSE}_{\mathbf{g}_{m,j}},
		\end{aligned}
	\end{equation}
	\begin{equation}\label{D2}
		\begin{aligned}
			&\mathbb{E}\{|\mathbf{u}_{m,p}^{\mathrm{H}}(\mathbf{H}_{m,l}-\hat{\mathbf{H}}_{m,l})\mathbf{x}_l|^2\}\\&=\frac{N\sigma^2\gamma_{m,l}}{\tau_{\mathrm{dl}} p_{\mathrm{dl}}\gamma_{m,l}+\sigma^2}\left(\displaystyle\sum_{k=1}^{K_{\mathrm{dl}}}\zeta_{l,k}^2p_{\mathrm{dl},l,k}\mathrm{Tr}(\mathbf{B}_{l,k})+\mathrm{Tr}(\mathbf{R}_{l})\right).
		\end{aligned}
	\end{equation}
	\indent The derivations of \eqref{D1} and \eqref{D2} are completed by using Lemma 1 for $\widetilde{\mathbf{g}}_{l,k}\sim \mathcal{CN}(\mathbf{0},\mathbf{I}_N)$ and $\mathbf{e}_{m,l}^{\mathrm{AP}}\sim \mathcal{CN}(\mathbf{0},\frac{\sigma^2\gamma_{m,l}}{\tau_{\mathrm{dl}} p_{\mathrm{dl}}\gamma_{m,l}+\sigma^2}\mathbf{I}_N)$.\\
	\indent For target sensing, as shown \eqref{sen}, the received signal strength after the receive filter consists of the desire echo signal $E_p^{\mathrm{ECHO}}$, the mutual interference $N_p^{\mathrm{MUI,S}}$ from the other targets, the residual interference $N_p^{\mathrm{MUI,S}}$ from the UL UEs and the interference $N_p^{\mathrm{D,S}}$ from the DL APs, which is calculated as follows
	\begin{equation}
		\begin{aligned}
			E_p^{\mathrm{ECHO}}&=\displaystyle\sum_{m=1}^{M}b_m^2\displaystyle\sum_{l=1}^{M}a_l^2\mathbb{E}\{|\mathbf{u}_{m,p}^{\mathrm{H}}\mathbf{G}_{m,p,l}\mathbf{x}_l|^2\}\\&\triangleq\displaystyle\sum_{m=1}^{M}\displaystyle\sum_{l=1}^{M}c_{\mathrm{s},m,p,l}^{(1)}a_l^2b_m^2,
		\end{aligned}
	\end{equation}
	\begin{equation}
		\begin{aligned}
			N_p^{\mathrm{MUI,S}}&=\displaystyle\sum_{m=1}^{M}b_m^2\displaystyle\sum_{l=1}^{M}a_l^2\displaystyle\sum_{t=1,t\neq p}^{K_{\mathrm{t}}}\mathbb{E}\{|\mathbf{u}_{m,p}^{\mathrm{H}}\mathbf{G}_{m,t,l}\mathbf{x}_l|^2\}\\&\triangleq\displaystyle\sum_{m=1}^{M}\displaystyle\sum_{l=1}^{M}c_{\mathrm{s},m,p,l}^{(\mathrm{I},1)}a_l^2b_m^2,
		\end{aligned}
	\end{equation}
	\begin{equation}
		\begin{aligned}
			N_p^{\mathrm{D,S}}&=\displaystyle\sum_{m=1}^{M}b_m^2\displaystyle\sum_{l=1}^{M}a_l^2\mathbb{E}\{|\mathbf{u}_{m,p}^{\mathrm{H}}(\mathbf{H}_{m,l}-\hat{\mathbf{H}}_{m,l})\mathbf{x}_l|^2\}\\&\triangleq\displaystyle\sum_{m=1}^{M}\displaystyle\sum_{l=1}^{M}c_{\mathrm{s},m,l}^{(\mathrm{I},2)}a_l^2b_m^2,
		\end{aligned}
	\end{equation}
	\begin{equation}
		\begin{aligned}
			N_p^{\mathrm{UL,C}}&=\displaystyle\sum_{m=1}^{M}b_m^2\displaystyle\sum_{j\in\mathcal{K}_{\mathrm{ul}}}\mathbb{E}\{|\mathbf{u}_{m,p}^{\mathrm{H}}\sqrt{p_{\mathrm{ul}}}(\mathbf{g}_{m,j}-\hat{\mathbf{g}}_{m,j})|^2\}\\&\triangleq\displaystyle\sum_{m=1}^{M}c_{\mathrm{s},m,p}^{(\mathrm{I},3)}b_m^2.
		\end{aligned}
	\end{equation}
	\indent By letting $c_{\mathrm{s},m,p,l}^{(2)}=c_{\mathrm{s},m,p,l}^{(\mathrm{I},1)}+c_{\mathrm{s},m,l}^{(\mathrm{I},2)}$ and $c_{\mathrm{s},m,p}^{(3)}=c_{\mathrm{s},m,p}^{(\mathrm{I},3)}+N\sigma^2$, the desired result in \eqref{sinr} is obtained.
	
	\bibliographystyle{IEEEtran}
	\bibliography{ref}

	% your own custom macro containing the $\backslash${\tt{includegraphics}} command to make things
	%\bf{If you include a photo:}\vspace{-33pt}
	%\begin{IEEEbiography}[{\includegraphics[width=1in,height=1.25in,clip,keepaspectratio]{fig1}}]{Michael Shell}
	%Use $\backslash${\tt{begin\{IEEEbiography\}}} and then for the 1st argument use $\backslash${\tt{includegraphics}} to declare and link the author photo.
	%Use the author name as the 3rd argument followed by the biography text.
	%\end{IEEEbiography}

	\vfill
	
\end{document}